\documentclass[journal]{IEEEtran}

\usepackage{amsmath,amssymb,amsfonts}
\usepackage{bm}
\usepackage{graphicx}
\usepackage{booktabs}
\usepackage{siunitx}

\usepackage[colorlinks=true, linkcolor=blue, citecolor=blue, urlcolor=blue]{hyperref}

\usepackage{mathtools}
\usepackage{microtype}

\usepackage{algorithm}
\usepackage{algpseudocode}
\usepackage{tabularx}
\usepackage{enumitem}
\usepackage{subcaption}

\usepackage{textcomp}
\usepackage{xcolor}
\usepackage{booktabs} 
\usepackage{multirow} 
\usepackage{array}    
\usepackage{tabularx}
\usepackage[section]{placeins} 

\usepackage[compress]{cite}

\begin{document}
	
	\title{A Heterogeneous Dual-Network Framework for Emergency Delivery UAVs: Communication Assurance and Path Planning Coordination}
	
	\author{Ping~Huang$^{1}$,
		Bin~Duo$^{1,*}$,~\IEEEmembership{Member,~IEEE,}
		Ziedor~Godfred$^{1}$,
		Liuwei~Huo$^{2}$,
		Jin~Ning$^{1}$,
		Xiaojun~Yuan$^{2,*}$,~\IEEEmembership{Fellow,~IEEE,}
		and~Jun~Li$^{3}$,~\IEEEmembership{Fellow,~IEEE,}
		\thanks{$^{1}$Ping Huang, Bin Duo, Ziedor Godfred, and Jin Ning are with the College of Computer Science and Cyber Security, Chengdu University of Technology, Chengdu 610059, China. $^{*}$Corresponding author (e-mail: duobin@cdut.edu.cn), xjyuan@uestc.edu.cn.}%
		\thanks{$^{2}$Liuwei Huo and Xiaojun Yuan are with the National Key Laboratory of Wireless Communications, University of Electronic Science and Technology of China, Chengdu 611731, China.}%
		\thanks{$^{3}$Jun Li is with the School of Information Science and Engineering, Southeast University, Nanjing 210096, China.}%
	}
	
	\markboth{Manuscript prepared for IEEE Transactions on Intelligent Transportation Systems}%
	{Huang \MakeLowercase{\textit{et al.}}: Resilient Heterogeneous UAV Logistics Network}
	
	\maketitle
	
	\begin{abstract}
		Natural disasters often damage ground infrastructure, making unmanned aerial vehicles (UAVs) essential for emergency supply delivery. Yet their safe operation in complex post-disaster environments requires reliable command-and-control (C2) links; link instability can cause loss of control, delay rescue, and trigger severe secondary harm. To provide continuous three-dimensional (3D) C2 coverage during dynamic missions, we propose a Heterogeneous Dual-Network Framework (HDNF) for safe and reliable emergency UAV delivery. HDNF tightly couples an Emergency Communication Support Network (ECSN), formed by hovering UAV base stations, with a Delivery Path Network (DPN), formed by fast-moving delivery UAVs. The ECSN dynamically ensures reliable communication for mission-critical flight corridors, while the DPN aligns trajectories with reliable coverage regions. We formulate a joint optimization problem over task assignment, 3D UAV-BS deployment, and path planning of DPN to maximize end-to-end C2 reliability while minimizing UAV flight energy consumption and base-station deployment cost. To solve this NP-hard problem, we develop a layered strategy with three components: (i) a multi-layer C2 service model that overcomes the limitations of traditional 2D metrics and aligns UAV-BS deployment with mission-critical 3D phases; (ii) a 3D coverage-aware multi-agent reinforcement learning algorithm that addresses the high-dimensional search space and improves both training efficiency and topology resilience; and (iii) a 3D communication-aware A* planner that jointly optimizes C2 quality and flight energy, mitigating trajectory--coverage mismatch and improving routing safety. Extensive simulations show that HDNF markedly improves C2 reliability, eliminates outages in critical phases, and sustains high task success rates while reducing UAV-BS deployment cost.
	\end{abstract}
	
	\section{Introduction}
	\IEEEPARstart{B}ig earthquakes and typhoons are becoming increasingly frequent and severe, straining emergency response systems \cite{background_1, background_2}. After such disasters, ground infrastructure is often badly damaged: roads may either be broken or blocked, and Ground Base Stations (BSs) frequently fail due to power outages or physical damage. This separates the damaged zones from the outside world, turning them into "information islands" \cite{background_3, UAV_BS_1}. Disaster relief protocols often emphasize the first 60 minutes after a disaster, usually termed as the "golden hour", during which swift delivery of aid can significantly reduce death rates. When ground transport is halted, unmanned aerial vehicles (UAVs) performing short-range airdrops become a key solution \cite{UAV_delivery_distence_1}.
	
	UAVs are widely used for disaster delivery because of rapid deployment and high mobility \cite{UAV_delivery_com_aware_1}. Prior studies optimize delivery performance through non-linear power control \cite{UAV_delivery_energy_1, UAV_delivery_energy_2}, payload constraints \cite{UAV_delivery_weight_1}, on-demand scheduling \cite{UAV_delivery_on_demand_1, UAV_delivery_on_demand_2}, and air-ground coordination with vehicles \cite{UAV_delivery_car_1, UAV_delivery_car_2}. However, these methods are mostly developed under favorable link conditions and typically rely on broad cellular coverage or stable Global Positioning System (GPS) links for navigation.
	
	In disaster zones, this ideal assumption often fails. When ground networks fail, steady links are not merely convenient; they are vital. Remotely operated UAVs rely on stable command and control (C2) links for telemetry, path updates, and safe obstacle avoidance \cite{UAV_delivery_com_aware_1}. Without steady links, UAVs lose timely guidance, which not only increases failure risks but also may threaten ground staff and property \cite{but}. Therefore, it is important to set up a strong communication network to deliver emergency supplies safely.
	
	Recent work uses UAV-BSs to rapidly restore connectivity in disaster areas through temporary aerial networks \cite{UAV_BS_1, UAV_BS_2}. Proper hovering altitude can improve LoS availability and reduce blockage \cite{UAV_BS_3}. However, most existing UAV-BS deployment methods target ground users \cite{background_3, UAV_BS_9}. When applied to moving delivery UAVs, base-station placement directly affects route feasibility: partial or static coverage can leave mission-critical phases (cargo drop, takeoff/landing, and cruise) exposed to C2 blind spots. Effective coordination between UAV-BSs and delivery UAVs in such dynamic scenarios remains underexplored.
	
	Most existing UAV-BS placement methods optimize service for fixed or slow-varying ground nodes (e.g., survivors or sensors) using fixed demand maps or slowly varying CSI \cite{UAV_BS_4, UAV_BS_5, UAV_BS_6, UAV_BS_7, UAV_BS_8, UAV_BS_10}. However, existing UAV-BS deployment schemes are primarily designed for 2D ground coverage, which inherently mismatches the 3D dynamic nature of delivery UAVs. Unlike static ground users, delivery UAVs require continuous C2 connectivity across varying altitudes, from vertical takeoff to high-altitude cruise. If UAV-BSs are deployed solely based on ground demand, severe communication blind spots will inevitably emerge along aerial corridors. Resolving this spatial mismatch requires tightly coupling the 3D placement of UAV-BSs with the trajectory planning of delivery UAVs. Unfortunately, this creates a highly complex, interdependent optimization problem. Conventional static-coverage optimization tools (e.g., MILP, convex optimization, and genetic algorithms) typically rely on fixed demand maps and adapt poorly to such a dynamically coupled, high-dimensional problem. Therefore, a joint 3D deployment-trajectory framework is imperative to ensure end-to-end C2 coverage.
	
	In this paper, we propose a Heterogeneous Dual-Network Framework (HDNF) that jointly designs an Emergency Communication Support Network (ECSN) and a Delivery Path Network (DPN). The ECSN consists of hovering UAV-BSs that provide a temporary C2 backbone, while the DPN consists of delivery UAVs executing 3D supply routes. Coordination is achieved by deploying ECSN nodes along mission corridors to establish reliable coverage regions, within which DPN trajectories are safely planned. We formulate a joint optimization problem over UAV-BS placement and delivery-UAV 3D path planning to maximize C2 reliability while minimizing flight energy and deployment cost under power and link constraints. To solve this coupled problem, we develop the following key components:
	
	\begin{itemize}
		\item[(i)] We propose a multi-layer C2 service model to address the limitation of traditional 2D coverage models in capturing the dynamic 3D communication requirements of delivery UAVs. The model maps spatial communication demand to distinct operational phases, including terminal supply delivery, vertical takeoff/landing, and high-speed cruise corridors. It provides a quantitative characterization of C2 link availability in 3D space and guides UAV-BS deployment to align with mission-critical flight phases rather than only ground coverage.
		\item[(ii)] We develop a 3D coverage-aware multi-agent twin delayed deep deterministic policy gradient algorithm with prioritized experience replay (3DCA-MATD3 with PER) to address the high-dimensional search space and computational bottleneck of UAV-BS placement in post-disaster scenarios. By using a shared backbone for global topological feature extraction and a prioritized experience replay mechanism for high-value sampling, the method improves training efficiency and deployment quality. In addition, by reformulating backhaul-connectivity evaluation with graph-theoretic metrics, the method avoids inefficient traversal-based checks and improves the resilience of the resulting UAV-BS topology.
		\item[(iii)] We develop a 3D communication-aware A* path-planning algorithm to reduce the risk of delivery UAVs entering communication dead zones. By integrating C2 link quality and flight energy consumption into a unified trajectory cost, the planner steers UAVs toward communication-reliable regions and mitigates trajectory--coverage mismatch, thereby improving routing safety under communication constraints.
	\end{itemize}
	
	Extensive simulations show that the proposed HDNF consistently outperforms representative baselines in both C2 link reliability and resource efficiency. The results confirm that 3D coverage-aware ECSN deployment effectively removes communication blind spots, enabling the DPN to generate safer and more energy-efficient 3D trajectories without excessive detours. Notably, this dual-network coordination satisfies stringent C2 requirements across all mission-critical phases, especially during the highly dynamic vertical takeoff/landing and high-altitude cruise segments. As the disaster area expands, HDNF maintains a higher task success rate with zero C2 outages while reducing the number of UAV-BS deployment requirements by up to 20\% compared with conventional static deployments.
	
	\begin{figure}[t]
		\centering
		\includegraphics[width=0.95\linewidth]{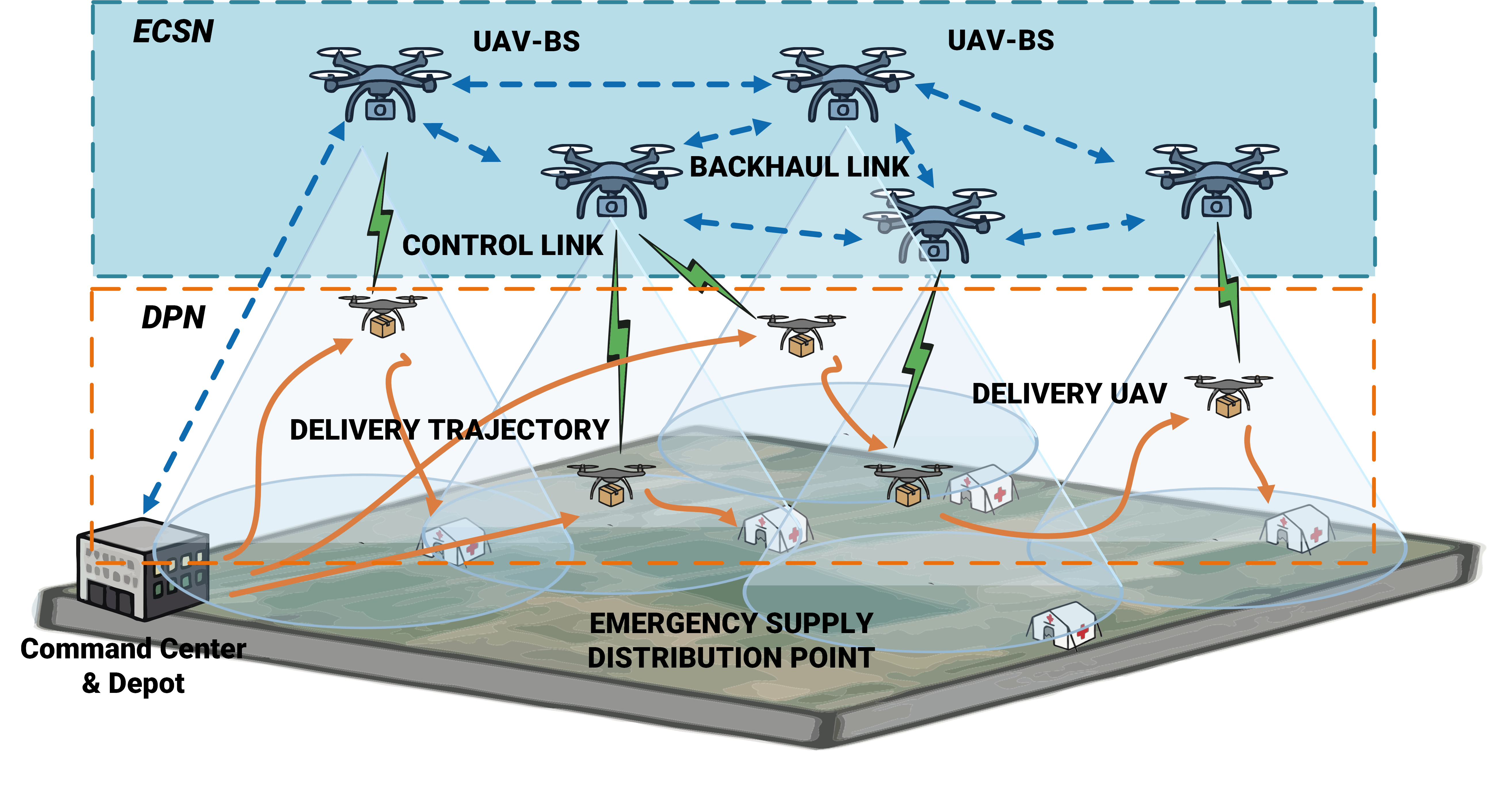}
		\caption{System model.}
		\label{fig:system_model}
	\end{figure}
	
	\section{System Model for HDNF}
	\label{sec:model}
	
	This section presents the system model of the proposed HDNF. The framework coordinates the ECSN and DPN to enable emergency supply delivery in disaster scenarios. In our model, the ECSN deploys UAV-BSs to provide temporary C2 connectivity, while the DPN plans delivery-UAV schedules and routes for emergency-supply requests under such connectivity support.
	
	As illustrated in Fig.~\ref{fig:system_model}, we study emergency supply delivery over a disaster region $\Omega \subset \mathbb{R}^3$. Delivery destinations are denoted by $\mathcal{T} = \{1, 2, \dots, T_{\text{task}}\}$, where $T_{\text{task}}$ is the number of tasks. The system adopts a heterogeneous architecture of delivery UAVs and UAV-BSs. Delivery UAVs, indexed by $\mathcal{U}_D = \{1, 2, \dots, U_D\}$, depart from a central depot and follow planned routes to serve $\mathcal{T}$. Each UAV-BS $b \in \mathcal{U}_B$ is fixed at $\mathbf{q}_b = (x_b, y_b, h_b)$ with altitude constraint $h_b \in [h_{\min}, h_{\max}]$. The ECSN aims to build a resilient aerial-to-aerial (A2A) mesh that maintains reliable C2 links for delivery UAVs throughout the disaster area.
	
	\subsection{ECSN Model}
	\label{set:UAVBS}
	Given the complex post-disaster environment, A2A links are vulnerable to blockage by high-rise obstacles. We therefore adopt a probabilistic channel model driven by LoS probability. For any aerial node pair (e.g., UAV-BS $b$ and delivery UAV $u$), the slot-$n$ 3D distance is $d_{b,u,n} = \|\mathbf{q}_b - \mathbf{p}_{u,n}\|$, where $\mathbf{p}_{u,n} = (x_{u,n}, y_{u,n}, z_{u,n})$ is the instantaneous position of delivery UAV $u$. Let $\theta_{b,u,n}$ denote their slot-$n$ elevation angle. In degrees, it is
	\begin{equation}
		\theta_{b,u,n} = \frac{180}{\pi} \arcsin \left( \frac{|h_b - z_{u,n}|}{d_{b,u,n}} \right).
		\label{eq:elevation}
	\end{equation}
	
	The LoS probability $P_{\text{LoS}}$ is modeled as a sigmoid of elevation angle:
	\begin{equation}
		P_{\text{LoS}}(\theta_{b,u,n}) = \frac{1}{1 + \alpha \exp \big( -\beta (\theta_{b,u,n} - \alpha) \big) },
		\label{eq:prob_los}
	\end{equation}
	where $\alpha$ and $\beta$ are environment-dependent constants reflecting the density of obstacles.
	
	The total path loss includes free-space path loss (FSPL) and environment-induced excess attenuation. FSPL in decibels (dB) is
	\begin{equation}
		\begin{split}
			L_{\text{FSPL}}(d_{b,u,n}) &= 20 \log_{10}(d_{b,u,n}) + 20 \log_{10}(f_c^{\text{(Hz)}}) \\
			&\quad + 20 \log_{10}\left(\frac{4\pi}{c_0}\right),
		\end{split}
		\label{eq:fspl}
	\end{equation}
	where $f_c^{\text{(Hz)}}$ is carrier frequency (Hz) and $c_0$ is the speed of light. If frequency is given in GHz, we convert it by $f_c^{\text{(Hz)}} = f_c^{\text{(GHz)}}\times 10^9$.
	
	Using channel statistics, the slot-$n$ average path loss $\bar{L}_{b,u,n}$ is written as the weighted sum of LoS and non-line-of-sight (NLoS) terms:
	\begin{equation}
		\begin{split}
			\bar{L}_{b,u,n} &= L_{\text{FSPL}}(d_{b,u,n}) + P_{\text{LoS}}(\theta_{b,u,n}) \eta_{\text{LoS}} \\
			&\quad + \big[1 - P_{\text{LoS}}(\theta_{b,u,n})\big] \eta_{\text{NLoS}},
		\end{split}
		\label{eq:avg_pl}
	\end{equation}
	where $\eta_{\text{LoS}}$ and $\eta_{\text{NLoS}}$ are the mean additional losses (dB) for LoS and NLoS links, respectively. The slot-$n$ average received power from $b$ to $u$ is
	\begin{equation}
		P_{b,u,n} = P_t^{\text{(lin)}} \cdot 10^{-\bar{L}_{b,u,n}/10},
		\label{eq:rx_power}
	\end{equation}
	where $P_t^{\text{(lin)}} = 10^{(P_t^{\text{(dBm)}}-30)/10}$ is the UAV-BS transmit power in Watts. The aggregated interference at delivery UAV $u$ in slot $n$ is
	\begin{equation}
		I_{u,n} = \sum_{b' \in \mathcal{U}_B \setminus \{b\}} P_{b',u,n},
		\label{eq:interference}
	\end{equation}
	and the corresponding signal-to-interference-plus-noise ratio (SINR) is
	\begin{equation}
		\mathrm{SINR}_{b,u,n} = \frac{P_{b,u,n}}{N_0^{\text{(lin)}} B + I_{u,n}},
		\label{eq:sinr}
	\end{equation}
	where $N_0^{\text{(lin)}}=10^{(N_0^{\text{(dBm/Hz)}}-30)/10}$ is noise power spectral density (W/Hz), and $B$ is system bandwidth. If a SINR threshold is specified in dB, we convert it as $\Gamma^{\text{(lin)}} = 10^{\Gamma^{\text{(dB)}}/10}$.
	
	To characterize ECSN topology, we model slot-$n$ backhaul as an undirected graph $\mathcal{G}_n=(\mathcal{V},\mathcal{E}_n)$, where $\mathcal{V}=\mathcal{U}_B\cup\{0\}$ includes all UAV-BSs and the central depot (node $0$). For any distinct $j,k\in\mathcal{V}$, $(j,k)$ is a candidate backhaul link. Using the same propagation/interference model as Eq.~(\ref{eq:sinr}), we compute directional qualities $\mathrm{SINR}_{j\to k,n}$ and $\mathrm{SINR}_{k\to j,n}$ and enforce reciprocity: $(j,k)$ is feasible only if both directions satisfy $\Gamma_{\text{bh}}$, i.e., $\min(\mathrm{SINR}_{j\to k,n},\mathrm{SINR}_{k\to j,n})\ge\Gamma_{\text{bh}}$.
	
	Accordingly, connectivity is encoded by adjacency matrix $\mathbf{A}_n \in \{0,1\}^{|\mathcal{V}| \times |\mathcal{V}|}$, whose entries are
	\begin{equation}
		a_{j,k,n} =
		\begin{cases}
			1, & \text{if } \substack{\min\!\left(\mathrm{SINR}_{j\to k,n},\mathrm{SINR}_{k\to j,n}\right) \geq \Gamma_{\text{bh}}, j \neq k}, \\
			0, & \text{otherwise}.
		\end{cases}
		\label{eq:adjacency}
	\end{equation}
	
	To quantify local connectivity, we compute node degree. For each $j \in \mathcal{V}$ in slot $n$, degree $d_{j,n}$ equals the number of feasible incident backhaul links:
	\begin{equation}
		d_{j,n} = \sum_{k \in \mathcal{V}} a_{j,k,n}.
		\label{eq:node_degree}
	\end{equation}
	This gives the diagonal degree matrix $\mathbf{D}_n = \text{diag}(d_{1,n}, d_{2,n}, \dots, d_{|\mathcal{V}|,n})$.
	
	We then construct Laplacian matrix $\mathbf{L}_n \in \mathbb{Z}^{|\mathcal{V}| \times |\mathcal{V}|}$ as $\mathbf{L}_n = \mathbf{D}_n - \mathbf{A}_n$, with entries
	\begin{equation}
		l_{j,k,n} =
		\begin{cases}
			d_{j,n}, & \text{if } j = k, \\
			-1, & \text{if } j \neq k \text{ and } a_{j,k,n} = 1, \\
			0, & \text{otherwise}.
		\end{cases}
		\label{eq:laplacian_construction}
	\end{equation}
	
	\begin{figure}[htbp]
		\centering
		\includegraphics[width=0.8\linewidth]{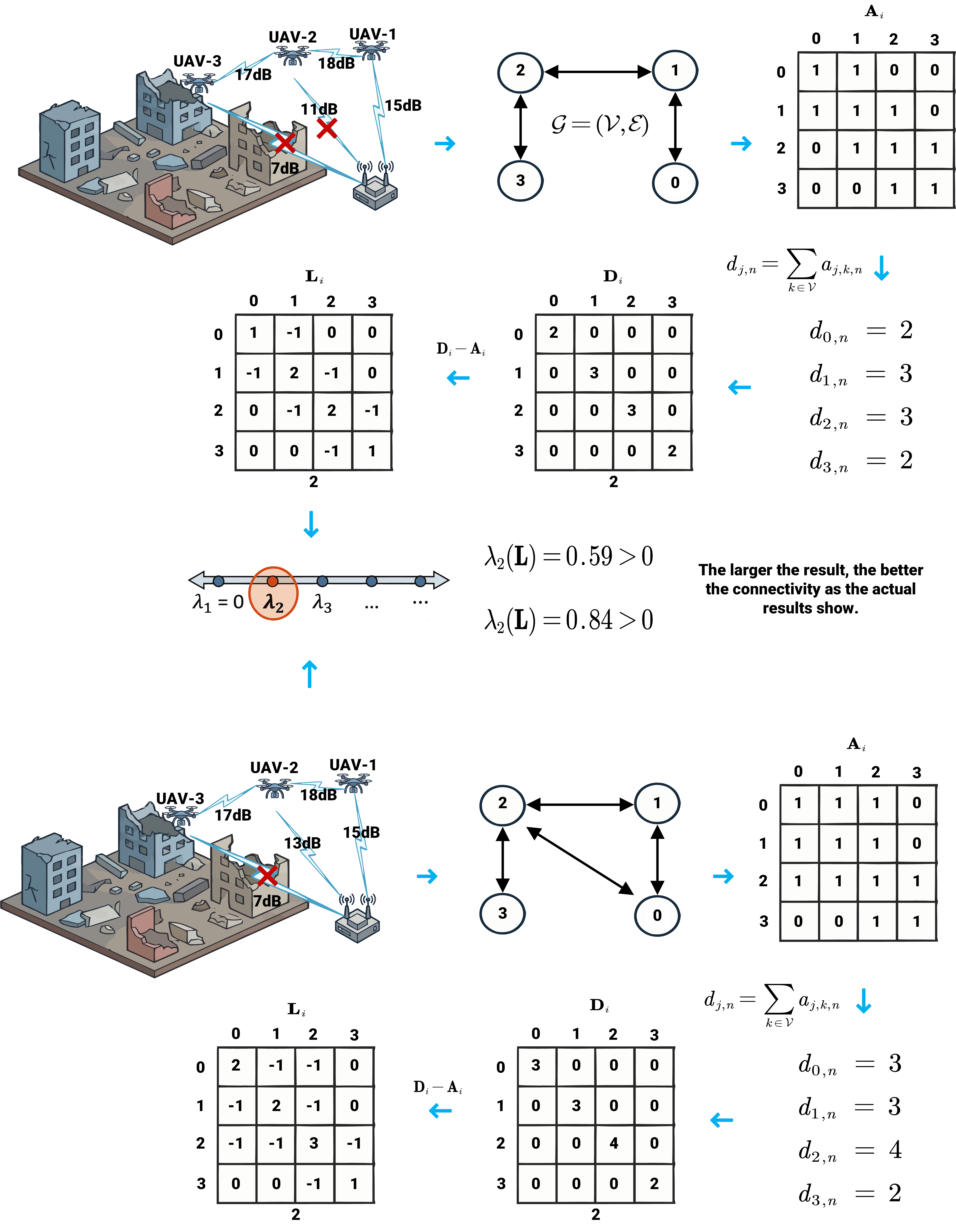}
		\caption{Comparison of different network topologies. A higher algebraic connectivity indicates a robust mesh network rather than a fragile chain.}
		\label{fig:connectivity_illustration}
	\end{figure}
	
	As illustrated in Fig.~\ref{fig:connectivity_illustration}, binary connectivity (path existence only) is insufficient for safety-critical disaster missions; structural robustness to link failures is also required. We therefore use algebraic connectivity, i.e., the second-smallest eigenvalue of the Laplacian, denoted by $\lambda_2(\mathbf{L}_n)$.
	
	From spectral graph theory, $\lambda_2(\mathbf{L}_n) > 0$ guarantees full connectivity. Beyond feasibility, a larger $\lambda_2(\mathbf{L}_n)$ indicates stronger robustness and fewer topological bottlenecks. We thus define a continuous connectivity utility that captures both feasibility and robustness:
	\begin{equation}
		C_{\text{conn},n} = \min \left( 1, \frac{\lambda_2(\mathbf{L}_n)}{\lambda_{\text{req}}} \right),
		\label{eq:connectivity_score}
	\end{equation}
	where $\lambda_{\text{req}}$ is the target robustness level. Under deployment $\mathbf{Q}$, we use the worst-slot utility $C_{\text{conn}}(\mathbf{Q}) \triangleq \min_n C_{\text{conn},n}(\mathbf{Q})$, which drives optimization toward resilient mesh formation.
	
	\subsection{DPN Model}
	\label{set:DUAV}
	
	Each delivery UAV $u$ has a maximum flight speed $v_{\text{D}}^{\max}$, each delivery UAV $u$ has a maximum cruise altitude $h_{\text{cruise}}$, and a maximum payload capacity $W_u^{\max}$. For tractable modeling, the trajectory of UAV $u$ is discretized into waypoints. UAV $u$ serves a task subset $\mathcal{T}_u \subseteq \mathcal{T}$, and its trajectory is represented as
	\begin{equation}
		\mathcal{P}_u = \{ \mathbf{p}_{u,0}, \mathbf{p}_{u,1}, \dots, \mathbf{p}_{u,n_u} \}.
	\end{equation}
	where $n_u$ is the index of the final waypoint (i.e., the trajectory contains $n_u+1$ waypoints from index $0$ to $n_u$).
	
	The total accumulated flight distance of UAV $u$ is
	\begin{equation}
		L_u = \sum_{i=0}^{n_u - 1} \| \mathbf{p}_{u,i+1} - \mathbf{p}_{u,i} \|
	\end{equation}
	
	For delivery UAV $u$, cruise-phase energy consumption is modeled by a widely used linear form for tractable optimization \cite{liner}:
	\begin{equation}
		E_u = \eta L_u \sum_{t \in \mathcal{T}_u} w_t,
	\end{equation}
	where $\eta$ (J/(m$\cdot$kg)) denotes the average payload-aware energy coefficient and $w_t$ is the payload weight associated with task $t$.
	
	To ensure mission feasibility, each delivery UAV must satisfy
	\begin{equation}
		\label{eq:ew}
		E_u \leq E_u^{\max}, \quad \sum_{t \in \mathcal{T}_u} w_t \leq W_u^{\max},
	\end{equation}
	where $E_u^{\max}$ represents the onboard battery energy budget.
	
	Each task $t \in \mathcal{T}$ is described by horizontal coordinates $(x_t,y_t)\in\Omega\subset\mathbb{R}^2$, corresponding 3D coordinates $\boldsymbol{\ell}_t=(x_t,y_t,0)\in\mathbb{R}^3$, and a service window $[a_t, b_t]$, where $a_t$ and $b_t$ are the earliest and latest expected arrival times. For delivery UAV $u$, let $t_u \in \mathcal{T}_u$ denote a local task index and let $\tau_{u,t_u}$ be the arrival time at task $t_u$. Then,
	\begin{equation}
		a_{t_u} \leq \tau_{u,t_u} \leq b_{t_u}, \quad \forall t_u \in \mathcal{T}_u.
		\label{eq:time_window}
	\end{equation}
	
	Task assignment enforces that each task is served exactly once by one delivery UAV:
	\begin{equation}
		\sum_{u \in \mathcal{U}_D} x_{u,t} = 1, \quad \forall t \in \mathcal{T},
		\label{eq:task_once}
	\end{equation}
	where $x_{u,t} \in \{0,1\}$ is a binary decision variable; $x_{u,t}=1$ indicates that UAV $u$ is assigned to global task $t$, and $x_{u,t}=0$ otherwise.
	
	Finally, each delivery trajectory must start from and return to the depot:
	\begin{equation}
		\mathbf{p}_{u,0} = \mathbf{p}_{u,n_u} = \mathbf{p}_{\text{center}}, \quad \forall u \in \mathcal{U}_D,
		\label{eq:depot_constraint}
	\end{equation}
	where $\mathbf{p}_{\text{center}}$ represents the coordinates of the command center and depot, $\mathbf{p}_{\text{center}}=(x_c,y_c,0)$.
	
	\subsection{Multi-Layer C2 Service Model}
	\label{subsec:synergy_model}
	
	Conventional UAV-BS deployment is usually optimized by ground-centric 2D metrics, which mismatch the dynamic 3D C2 requirements of delivery UAVs. A complete mission includes terminal delivery, vertical takeoff/landing, and high-altitude cruise, and each phase has different C2 vulnerabilities. A single aggregated metric can therefore hide phase-specific blind spots. To address this, we build a multi-layer C2 service model that quantifies phase-dependent communication quality and aligns ECSN deployment with actual DPN trajectories.
	
	\begin{figure}[t]
		\centering
		\includegraphics[width=0.65\linewidth]{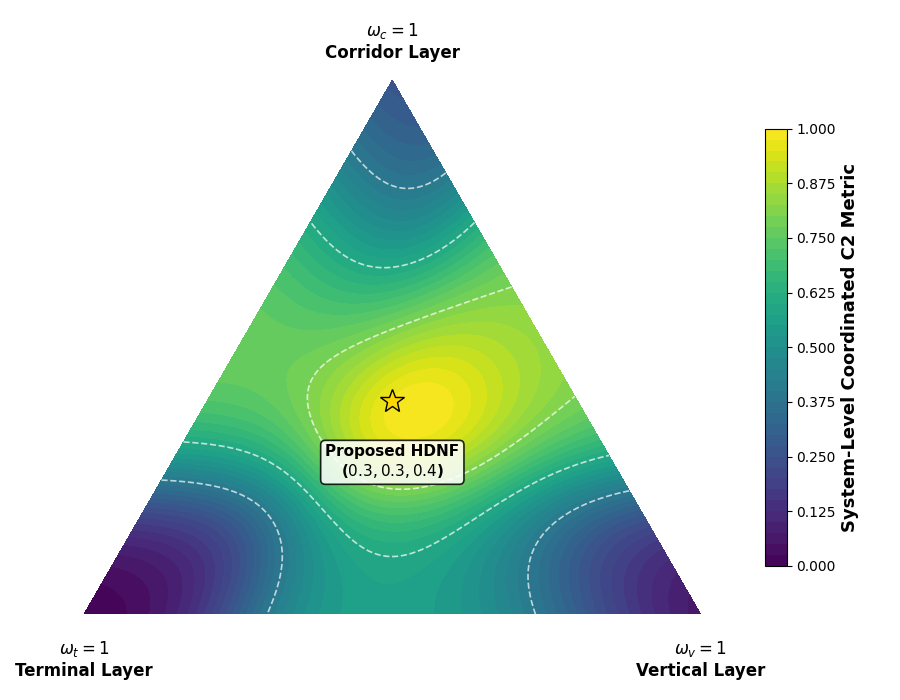}
		\caption{Phase transition of the system-level coordinated C2 metric under varying layer weights $(\omega_t, \omega_v, \omega_c)$.}
		\label{fig:weight_impact}
	\end{figure}
	
	For each sampled point $\mathfrak{x}\in\mathbb{R}^3$, we define a binary C2-feasibility indicator and a normalized Shannon-capacity metric:
	\begin{equation}
		\begin{aligned}
			\mathbb{I}(\mathfrak{x}) &= \mathbb{I}\left( \max_{b \in \mathcal{U}_B} \mathrm{SINR}(\mathfrak{x}, b) \ge \Gamma_{\text{ctrl}} \right), \\
			\tilde{C}(\mathfrak{x}) &= \min\left(1, \frac{\log_2\big(1 + \max_{b \in \mathcal{U}_B}\mathrm{SINR}(\mathfrak{x}, b)\big)}{C_{\max}}\right),
		\end{aligned}
		\label{eq:indicator_and_capacity}
	\end{equation}
	where $\Gamma_{\text{ctrl}}$ is the minimum C2 SINR threshold and $C_{\max}$ is a normalization constant for spectral efficiency. Eq.~(\ref{eq:indicator_and_capacity}) determines whether a point is C2-feasible and, if feasible, how strong its C2 quality is.
	
	To capture phase heterogeneity, we construct three sampling sets:
	\begin{enumerate}
		\item \textbf{Terminal layer} ($\mathcal{S}_t$):
		\begin{equation}
			\mathcal{S}_t = \{\boldsymbol{\ell}_t \mid t \in \mathcal{T}\},
		\end{equation}
		which contains all terminal service locations, $\boldsymbol{\ell}_t=(x_t,y_t,0)$.
		\item \textbf{Vertical layer} ($\mathcal{S}_v$): for each task location $\boldsymbol{\ell}_t$, vertical samples are
		\begin{equation}
			\begin{aligned}
				\mathfrak{x}_{t,m} &= (x_t,y_t,h_{t,m}),\\
				h_{t,m} &= m\Delta h,\quad m=0,\dots,M_v.
			\end{aligned}
		\end{equation}
		and $\mathcal{S}_v=\{\mathfrak{x}_{t,m}\}$.
		\item \textbf{Corridor layer} ($\mathcal{S}_f$): for each task $t$, we sample points on the nominal depot-to-task cruise corridor at altitude $h_{\text{cruise}}$:
		\begin{equation}
			\begin{aligned}
				\mathfrak{x}_{t,\iota}
				&=\left(x_c+\frac{\iota}{I_t}(x_t-x_c),\right.\\
				&\quad\left.y_c+\frac{\iota}{I_t}(y_t-y_c),\;h_{\text{cruise}}\right),\\
				&\quad \iota=0,\dots,I_t,
			\end{aligned}
		\end{equation}
		and $\mathcal{S}_f=\{\mathfrak{x}_{t,\iota}\}$.
	\end{enumerate}
	Here, $\Delta h$ is the vertical sampling interval, $M_v$ is the number of vertical sampling intervals per task, and $I_t$ controls corridor sampling density.
	
	Based on these sets, we compute the per-layer coverage ratio and average normalized capacity:
	\begin{equation}
		\begin{aligned}
			C_{\text{term}} &= \frac{1}{|\mathcal{S}_t|} \sum_{\mathfrak{x} \in \mathcal{S}_t} \mathbb{I}(\mathfrak{x}), \quad & \bar{C}_t &= \frac{1}{|\mathcal{S}_t|} \sum_{\mathfrak{x} \in \mathcal{S}_t} \tilde{C}(\mathfrak{x}), \\
			C_{\text{vert}} &= \frac{1}{|\mathcal{S}_v|} \sum_{\mathfrak{x} \in \mathcal{S}_v} \mathbb{I}(\mathfrak{x}), \quad & \bar{C}_v &= \frac{1}{|\mathcal{S}_v|} \sum_{\mathfrak{x} \in \mathcal{S}_v} \tilde{C}(\mathfrak{x}), \\
			C_{\text{corr}} &= \frac{1}{|\mathcal{S}_f|} \sum_{\mathfrak{x} \in \mathcal{S}_f} \mathbb{I}(\mathfrak{x}), \quad & \bar{C}_c &= \frac{1}{|\mathcal{S}_f|} \sum_{\mathfrak{x} \in \mathcal{S}_f} \tilde{C}(\mathfrak{x}).
		\end{aligned}
		\label{eq:layer_metrics}
	\end{equation}
	which respectively quantify the spatial availability and the link quality for each mission phase; specifically, the left column evaluates the proportion of locations satisfying the minimum C2 threshold as a measure of coverage reliability, while the right column measures the average spectral efficiency across the sampled regions to reflect the achievable data rate.
	
	We then define the system-level coordinated C2 service metric as
	\begin{equation}
		\bar{C}_{\text{syn}} = \omega_t \bar{C}_t + \omega_v \bar{C}_v + \omega_c \bar{C}_c, \quad (\omega_t + \omega_v + \omega_c = 1).
		\label{eq:avg_capacity_total}
	\end{equation}
	where $(\omega_t,\omega_v,\omega_c)$ are phase-importance weights.
	
	As illustrated in Fig.~\ref{fig:weight_impact}, the proposed coordinated C2 metric demonstrates distinct performance trade-offs under varying weight settings. The three vertices of the ternary plot represent extreme single-phase focus (e.g., prioritizing only terminal delivery or only cruise corridors), which inevitably compromises communication reliability in the neglected phases. In contrast, the balanced weight setting in HDNF effectively captures the heterogeneous 3D demands, thereby preventing phase-specific blind spots.
	
	Consequently, this multi-layer C2 model establishes a comprehensive evaluation standard for the dynamic 3D flight process. By utilizing $\bar{C}_{\text{syn}}$ as a unified metric, the framework shifts the evaluation focus away from redundant 2D ground coverage, explicitly ensuring robust, end-to-end C2 connectivity along the actual 3D flight corridors.
	
	\section{Problem Formulation}
	\label{sec:problem_formulation}
	
	In this section, the dual network coordination problem is formulated as a joint optimization problem of task assignment, ECSN deployment, and DPN trajectory planning. The objective is to solve the problem of interdependence between C2 communication coverage and flight trajectory feasibility. The problem is subject to the following three sets of decision variables:
	\begin{itemize}
		\item Task assignment $\mathbf{X}$: The binary assignment matrix $\mathbf{X}=\{x_{u,t}\} \in \{0,1\}^{U_D \times T_{\text{task}}}$, where each $x_{u,t} \in \{0,1\}$ is a scalar indicator denoting whether delivery UAV $u \in \mathcal{U}_D$ is assigned to task $t \in \mathcal{T}$.
		\item ECSN deployment $\mathbf{Q}$: The 3D coordinate set $\mathbf{Q} = \{\mathbf{q}_b \mid b \in \mathcal{U}_B\}$, determining the topology of the aerial C2 backbone.
		
		\item DPN trajectories $\boldsymbol{\mathcal{P}}$: The delivery trajectory set $\boldsymbol{\mathcal{P}} = \{\mathcal{P}_u \mid u \in \mathcal{U}_D\}$, representing the ordered waypoints of each delivery mission.
	\end{itemize}
	Therefore, we formulate the following optimization problem:
	\begin{equation}
		\begin{aligned}
			\min_{\mathbf{X}, \mathbf{Q}, \boldsymbol{\mathcal{P}}} \quad &
			\omega_d \,\operatorname{card}(\mathbf{Q}) + \omega_e \sum_{u \in \mathcal{U}_D} E_u \\
			&\quad - \left( \gamma_t C_{\text{term}} + \gamma_v C_{\text{vert}} + \gamma_c C_{\text{corr}} \right), \\[3pt]
			\text{s.t.} \quad &
			(\ref{eq:ew}),\; (\ref{eq:time_window}),\; (\ref{eq:task_once}),\; (\ref{eq:depot_constraint}), \quad \\
			& C_{\text{conn}}(\mathbf{Q}) \geq 0.5, \quad \\
			& h_{\min} \le h_b \le h_{\max}, \quad \forall \mathbf{q}_b \in \mathbf{Q}
		\end{aligned}
		\label{eq:optimization}
	\end{equation}
	$\operatorname{card}(\mathbf{Q})$ denotes the number of deployed UAV-BSs. The coefficients $\omega_d$ and $\omega_e$ weight the deployment cost and the aggregate flight energy $\sum_{u \in \mathcal{U}_D} E_u$, respectively. $\gamma_t$, $\gamma_v$, and $\gamma_c$ are the phase-priority weights assigned to the terminal, vertical, and corridor C2 coverage metrics, respectively.
	
	Problem (\ref{eq:optimization}) is a mixed-integer nonlinear program (MINLP) and is NP-hard. Its main difficulty is variable coupling: $\mathbf{Q}$ must match the spatiotemporal demand induced by $\boldsymbol{\mathcal{P}}$, while trajectory feasibility depends on the C2 coverage provided by $\mathbf{Q}$. The nonconvex A2A channel model and discrete structure of $\mathbf{X}$ further make conventional optimization computationally prohibitive. We therefore adopt a MARL-driven hierarchical decomposition to decouple variables and efficiently explore the high-dimensional mixed-variable space (Section~\ref{sec:method}).
	
	\section{MARL-Based Hierarchical Decomposition Scheme}
	\label{sec:method}
	
\begin{figure}[!t]
		\centering
		\includegraphics[width=\linewidth]{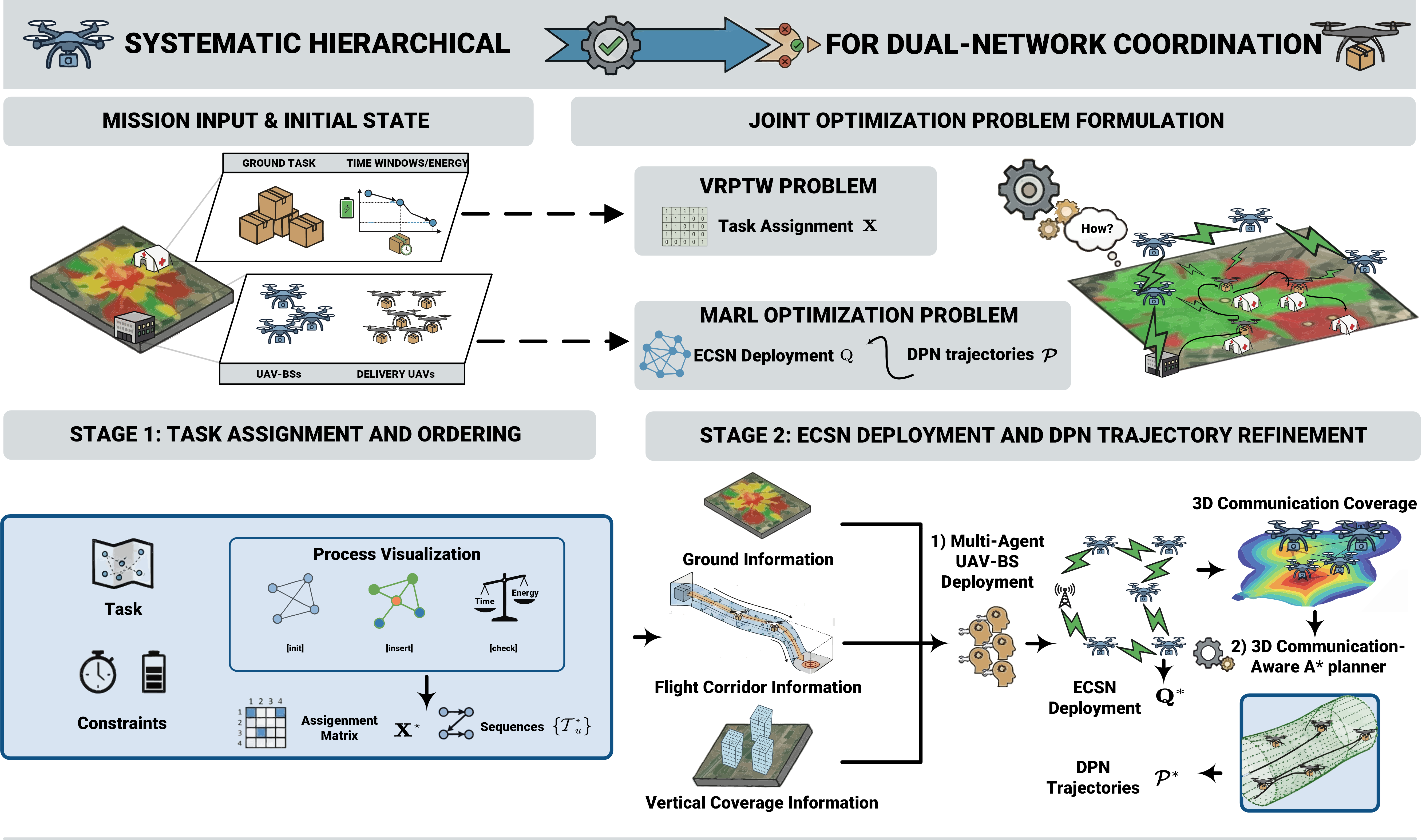}
		\caption{Proposed HDNF workflow.}
		\label{fig:framework}
	\end{figure}

	This section presents the MARL-based hierarchical decomposition scheme for Problem~(\ref{eq:optimization}), as illustrated in Fig.~\ref{fig:framework}. Specifically, we hierarchically decompose the original problem into two coordinated sub-problems: 1) a task-assignment and ordering sub-problem, and 2) an ECSN deployment and DPN trajectory-refinement sub-problem.
	
	\subsection{Problem Division}
	\label{sec:framework}
	
	When the task sequence assigned to delivery UAV $u$ is denoted by $\mathcal{T}_u = \{t_{u,1}, \dots, t_{u,|\mathcal{T}_u|}\}$, the corresponding flight distance is calculated as follows:
	\begin{equation}
		\begin{split}
			\tilde{L}_u &= \|\mathbf{p}_{\text{center}} - \boldsymbol{\ell}_{t_{u,1}}\| \\
			&\quad + \sum_{k=1}^{|\mathcal{T}_u|-1} \|\boldsymbol{\ell}_{t_{u,k+1}} - \boldsymbol{\ell}_{t_{u,k}}\| \\
			&\quad + \|\boldsymbol{\ell}_{t_{u,|\mathcal{T}_u|}} - \mathbf{p}_{\text{center}}\|.
		\end{split}
		\label{eq:estimated_length}
	\end{equation}
	Correspondingly, its flight energy consumption is $\tilde{E}_u = \eta \tilde{L}_u\sum_{t\in\mathcal{T}_u} w_t$.
	
	Based on the above derivation, we formulate the delivery-UAV task assignment and ordering sub-problem as follows:
	\begin{equation}
		\begin{aligned}
			\min_{\mathbf{X}, \{\mathcal{T}_u\}_{u\in\mathcal{U}_D}} \quad
			& \omega_e \sum_{u \in \mathcal{U}_D} \tilde{E}_u \\[3pt]
			\text{s.t.} \quad
			& \tilde{E}_u \le E_u^{\max}, \quad \forall u \in \mathcal{U}_D,\\[2pt]
			& (\ref{eq:ew}),\; (\ref{eq:time_window}),\; (\ref{eq:task_once}),\; (\ref{eq:depot_constraint}), \quad \\
			& x_{u,t} \in \{0,1\}, \quad \forall u \in \mathcal{U}_D,\; \forall t \in \mathcal{T}.
		\end{aligned}
		\label{eq:UL_final}
	\end{equation}
	
	Given $\mathbf{X}^{\ast}$ and $\{\mathcal{T}_u^{\ast}\}_{u\in\mathcal{U}_D}$, the second sub-problem for ECSN deployment and DPN trajectory refinement is formulated as:
	\begin{equation}
		\begin{aligned}
			\min_{\boldsymbol{\mathcal{P}}, \mathbf{Q}} \quad
			& \omega_d \,\operatorname{card}(\mathbf{Q}) + \omega_e \sum_{u \in \mathcal{U}_D} E_u(\mathcal{P}_u) \\
			& - \left(\gamma_t C_{\text{term}} + \gamma_v C_{\text{vert}} + \gamma_c C_{\text{corr}}\right), \\[3pt]
			\text{s.t.} \quad
			& E_u(\mathcal{P}_u) \le E_u^{\max}, \quad \forall u \in \mathcal{U}_D,\\[2pt]
			& C_{\text{conn}}(\mathbf{Q}) \geq 0.5, \\[2pt]
			& h_{\min} \le h_b \le h_{\max}, \quad \forall \mathbf{q}_b \in \mathbf{Q}. \\[2pt]
			& (\ref{eq:depot_constraint}).
		\end{aligned}
		\label{eq:LL_final}
	\end{equation}
	\subsection{Stage One: Task Assignment and Ordering}
	
	For solving Problem~(\ref{eq:UL_final}), we propose a sequential insertion algorithm to rapidly obtain sub-optimal solutions for task assignments.
	
	The core idea is to process tasks sequentially one after the other rather than evaluating them through numerous sorting combinations. Therefore, we initialize the sorting based on the urgency level of each task (according to the time window). This yields a task sequence with urgency levels.
	
	Next, we define the following marginal cost to evaluate which delivery UAV should be assigned to which insertion position $k$ in the current route:
	\begin{equation}
		\Delta J_{u,t,\mu} = \omega_e \Delta \tilde{E}_{u,t} + \omega_{\text{wait}} \Delta W_{u,t},
		\label{eq:incremental_cost}
	\end{equation}
	where $\Delta \tilde{E}_{u,t}=\tilde{E}_u^{\text{after}}-\tilde{E}_u^{\text{before}}$ denotes the exact incremental payload-aware energy, with
	$\tilde{E}_u^{\text{before}}=\eta\tilde{L}_u^{\text{before}}\sum_{i\in\mathcal{T}_u^{\text{before}}}w_i$ and
	$\tilde{E}_u^{\text{after}}=\eta\tilde{L}_u^{\text{after}}\sum_{i\in\mathcal{T}_u^{\text{after}}}w_i$.
	Here, $\omega_{\text{wait}}$ is the waiting-time penalty weight in the insertion cost.
	\begin{equation}
		\Delta W_{u,t} = \max(0, a_t - \tau_{u,t}),
		\label{eq:incremental_wait}
	\end{equation}
	which represents the incremental waiting-time penalty for premature arrivals. If the UAV arrives earlier than the earliest allowable service time $a_t$, it must hover and wait.
	
	Considering only marginal cost is insufficient. We must also determine whether the current insertion would violate energy or time-window constraints. If either occurs, we set $\Delta J_{u,t,\mu}=\mathcal{M}_{\infty}$, where $\mathcal{M}_{\infty}$ is a sufficiently large infeasibility penalty constant.
	
	Finally, after traversing all delivery UAVs and all valid insertion positions $\mu$, we select the combination $(u^*,\mu^*)$ with the minimum total cost. This combination is added to the UAV's task sequence $\mathcal{T}_u^{\ast}$, and the solution variables $\mathbf{X}^*$ are updated. If no insertion is feasible for the current task, the algorithm terminates and reports the instance as infeasible, which is consistent with Eq.~(\ref{eq:task_once}). The pseudocode is shown in Algorithm~\ref{alg:heuristic}.
	
	\begin{algorithm}[t]
		\caption{Sequential Insertion Heuristic for Task Assignment}
		\label{alg:heuristic}
		\begin{algorithmic}[1]
			\State \textbf{Input:} Task set $\mathcal{T}$, delivery-UAV set $\mathcal{U}_D$.
			\State \textbf{Initialize:} $\mathcal{T}_u^{\ast} \leftarrow \emptyset$ and $\mathbf{X}^{\ast} \leftarrow \mathbf{0}$ for all $u \in \mathcal{U}_D$.
			\State Sort tasks by urgency to form $\mathcal{T}_{\text{rem}} \leftarrow \mathcal{T}$ (according to time windows).
			\While{$\mathcal{T}_{\text{rem}}$ is not empty}
			\State $t \leftarrow \text{Dequeue}(\mathcal{T}_{\text{rem}})$
			\State $J_{\min} \leftarrow \mathcal{M}_{\infty}$; $(u^{\ast},\mu^{\ast}) \leftarrow \emptyset$
			\ForAll{$u \in \mathcal{U}_D$ and valid insertion positions $\mu$ in route $\mathcal{T}_u^{\ast}$}
			\State Compute $\Delta J_{u,t,\mu}$ by Eq.~\eqref{eq:incremental_cost}
			\State \textbf{if} energy/time-window constraints are violated, set $\Delta J_{u,t,\mu} \leftarrow \mathcal{M}_{\infty}$
			\If{$\Delta J_{u,t,\mu} < J_{\min}$}
			\State $J_{\min} \leftarrow \Delta J_{u,t,\mu}$; $(u^{\ast},\mu^{\ast}) \leftarrow (u,\mu)$
			\EndIf
			\EndFor
			\If{$J_{\min} < \mathcal{M}_{\infty}$}
			\State Insert task $t$ into $\mathcal{T}_{u^{\ast}}^{\ast}$ at position $k^{\ast}$
			\State Update assignment matrix $\mathbf{X}^{\ast}$
			\Else
			\State \textbf{Return} infeasible (no feasible insertion exists for task $t$)
			\EndIf
			\EndWhile
			\State \textbf{Return} sub-optimal solution $(\mathbf{X}^{\ast}, \{\mathcal{T}_u^{\ast}\}_{u\in\mathcal{U}_D})$.
		\end{algorithmic}
	\end{algorithm}
	
	For compactness in the hierarchical pipeline, we denote the Stage-One routine in Algorithm~\ref{alg:heuristic} by the following abstract interface $\text{Heuristic\_VRPTW}(\mathcal{T},\mathcal{U}_D,\omega_e,\omega_{\text{wait}})=(\mathbf{X}^{\ast}, \{\mathcal{T}_u^{\ast}\}_{u\in\mathcal{U}_D})$.
	\subsection{Stage Two: ECSN Deployment and DPN Trajectory Refinement}
	\label{subsec:stage_two}
	
	To efficiently solve the sub-problem formulated in Eq.~(\ref{eq:LL_final}), this stage is divided into two sequential steps, achieving an effective hierarchical decomposition of ECSN deployment and DPN trajectory planning. Specifically, we first propose a 3D-CASB MATD3 with PER to optimize the spatial topology of the UAV-BSs. Subsequently, based on the established communication coverage, we introduce a 3D communication-aware A* planner to refine the flight trajectories of the delivery UAVs. This decoupling approach ensures effective coordination between the ECSN and the DPN.
	
	\subsubsection{3D-CASB MATD3 with PER for ECSN Deployment}
	
	We reformulate the UAV-BS deployment sub-problem in~(\ref{eq:LL_final}) as a Multi-Agent Markov Decision Process (MMDP), where nonconvex continuous 3D deployment is handled through sequential cooperative decisions under centralized training and decentralized execution (CTDE).
	
	\textbf{a) Multi-layer grid-map state construction:} We build the communication-demand map from the multi-layer C2 model in Section~\ref{subsec:synergy_model}. Samples in $\mathcal{S}_t$, $\mathcal{S}_f$, and $\mathcal{S}_v$ are projected to a structured grid (rather than an unstructured 3D point list), which stabilizes policy learning (Fig.~\ref{fig:framework}). Let the operational airspace be $\Omega\subset\mathbb{R}^3$ with horizontal projection $\Omega_{xy}\subset\mathbb{R}^2$. We uniformly partition $\Omega_{xy}$ into $K\times K$ cells and define
	\begin{equation}
		M_{\text{grid}}\in[0,1]^{C\times K\times K},\qquad C=3,
	\end{equation}
	where the three channels correspond to terminal service demand ($c=1$), cruise-corridor demand ($c=2$), and vertical takeoff/landing demand ($c=3$), i.e., $(\mathcal{S}^{(1)},\mathcal{S}^{(2)},\mathcal{S}^{(3)})=(\mathcal{S}_t,\mathcal{S}_f,\mathcal{S}_v)$.
	
	For each channel $c\in\{1,2,3\}$ with sampled set $\mathcal{S}^{(c)}$, and each grid index pair $(\mathfrak{r},\mathfrak{c})\in\{1,\ldots,K\}^2$, let $\mathrm{cell}(\mathfrak{r},\mathfrak{c})\subset\Omega_{xy}$ denote the corresponding grid cell and define the in-cell set
	\begin{equation}
		\mathcal{K}_{\mathfrak{r},\mathfrak{c}}^{(c)}=\mathcal{S}^{(c)}\cap\mathrm{cell}(\mathfrak{r},\mathfrak{c}).
	\end{equation}
	Using the C2-availability indicator $\mathbb{I}(\cdot)$ in Eq.~(\ref{eq:indicator_and_capacity}), the in-cell coverage ratio is
	\begin{equation}
		\kappa_{\mathfrak{r},\mathfrak{c}}^{(c)}=
		\begin{cases}
			\dfrac{1}{\left|\mathcal{K}_{\mathfrak{r},\mathfrak{c}}^{(c)}\right|}\sum\limits_{p\in\mathcal{K}_{\mathfrak{r},\mathfrak{c}}^{(c)}}\mathbb{I}(p), & \left|\mathcal{K}_{\mathfrak{r},\mathfrak{c}}^{(c)}\right|>0, \\
			1, & \text{otherwise}.
		\end{cases}
	\end{equation}
	The corresponding outage-density value is
	\begin{equation}
		\label{eq:grid_map}
		M_{\text{grid}}^{(c)}[\mathfrak{r},\mathfrak{c}]=1-\kappa_{\mathfrak{r},\mathfrak{c}}^{(c)}.
	\end{equation}
	Thus, each grid cell is represented by
	\begin{equation}
		\mathrm{cell}(\mathfrak{r},\mathfrak{c})=\big[M_{\text{grid}}^{(1)}[\mathfrak{r},\mathfrak{c}],\,M_{\text{grid}}^{(2)}[\mathfrak{r},\mathfrak{c}],\,M_{\text{grid}}^{(3)}[\mathfrak{r},\mathfrak{c}]\big]^\top\in[0,1]^3,
	\end{equation}
	where larger components indicate more uncovered demand in the corresponding layer.
	
	\begin{figure*}[t]
		\centering
		\includegraphics[width=0.8\linewidth]{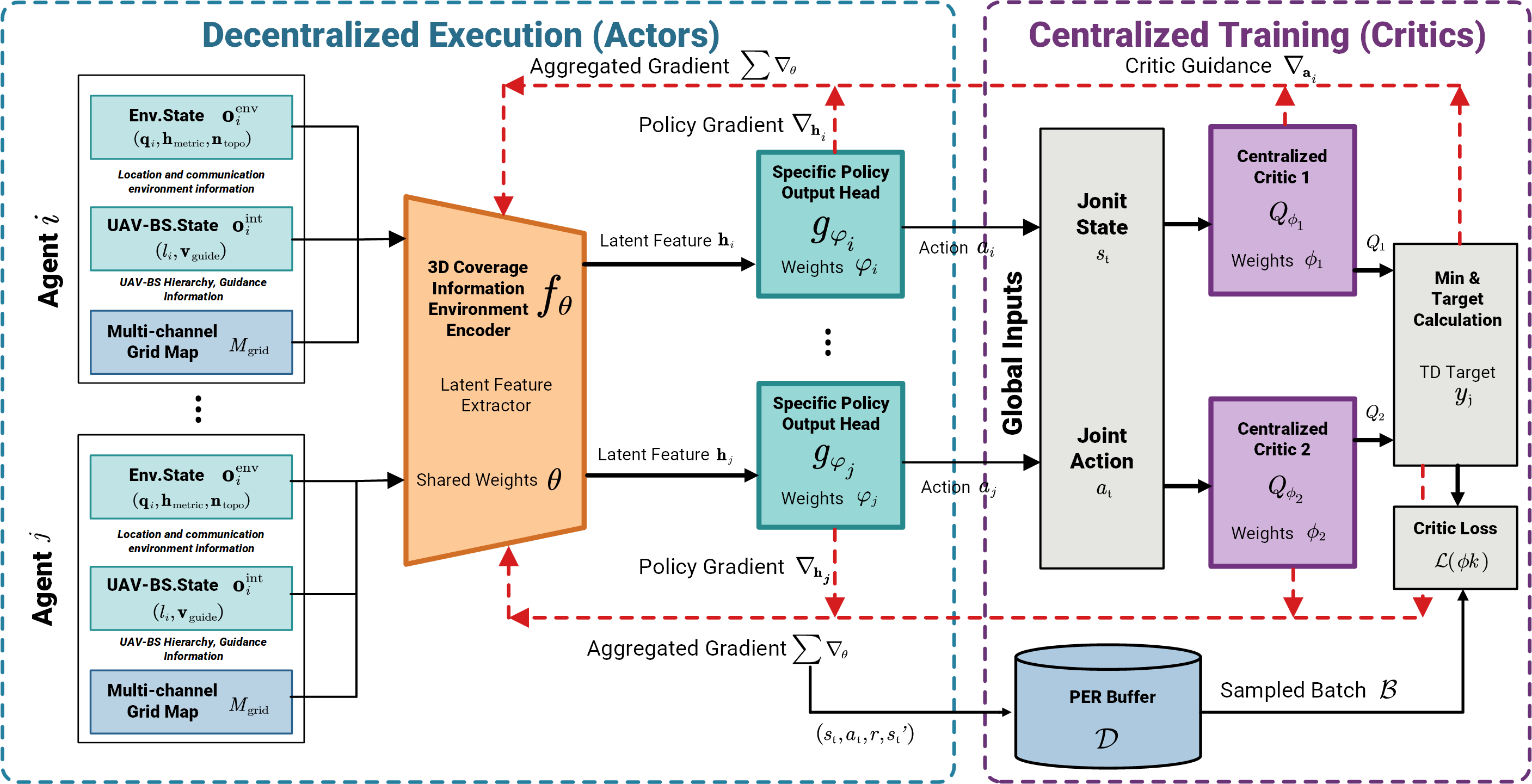}
		\caption{Pre-trained 3D-CASB-MATD3 with PER architecture.}
		\label{fig:matd3_arch}
	\end{figure*}
	
	\textbf{b) Agent observation, action, and deployment set:} Agent $i$ observes
	\begin{equation}
		\mathbf{o}_i=\{\mathbf{o}^{\text{env}}_i,\mathbf{o}^{\text{int}}_i\},
	\end{equation}
	with
	\begin{equation}
		\mathbf{o}^{\text{env}}_i=\{\mathbf{q}_i,\mathbf{h}_{\text{metric}},\mathbf{n}_{\text{topo}}\},
	\end{equation}
	\begin{equation}
		\mathbf{o}^{\text{int}}_i=\{l_i,\mathbf{v}_{\text{guide}}\}.
	\end{equation}
	Here, $\mathbf{q}_i\in\mathbb{R}^3$ is the normalized agent position, $\mathbf{h}_{\text{metric}}=[C_{\text{term}},C_{\text{vert}},C_{\text{corr}},C_{\text{conn}}]$ is the global scalar-metric vector, $\mathbf{n}_{\text{topo}}$ stacks relative-position vectors of the $k$ nearest neighbors, $l_i\in\{0,1\}$ is the altitude-layer indicator, and $\mathbf{v}_{\text{guide}}\in\mathbb{R}^3$ points toward uncovered-demand directions from $M_{\text{grid}}$.
	
	At MARL decision step $\mathfrak{t}\in\{0,\dots,T_{\text{step}}\}$, agent $i$ outputs
	\begin{equation}
		\mathbf{a}_i^{(\mathfrak{t})}=\big[\mathbf{v}_{\text{cmd},i}^{(\mathfrak{t})},\,g_{\text{gate},i}^{(\mathfrak{t})}\big],
	\end{equation}
	where $\mathbf{v}_{\text{cmd},i}^{(\mathfrak{t})}\in[-1,1]^3$ is a normalized 3D motion command used to update $\mathbf{q}_i$, and $g_{\text{gate},i}^{(\mathfrak{t})}\in[0,1]$ is a deployment-gate score that determines whether the UAV-BS candidate is retained. The position update is
	\begin{equation}
		\mathbf{q}_i^{(\mathfrak{t}+1)}=\mathrm{Clip}_{\Omega_{xy}\times[h_{\min},h_{\max}]}\!\left(\mathbf{q}_i^{(\mathfrak{t})}+v_{\text{D}}^{\max}\mathbf{v}_{\text{cmd},i}^{(\mathfrak{t})}\Delta t\right),
		\label{eq:kinematic_update}
	\end{equation}
	where $\mathrm{Clip}_{\Omega_{xy}\times[h_{\min},h_{\max}]}(\cdot)$ clips horizontal coordinates into $\Omega_{xy}$ and altitude into $[h_{\min},h_{\max}]$, and $\Delta t>0$ denotes the physical time interval of one MARL decision step. Agent $i$ is activated if $g_{\text{gate},i}^{(T_{\text{step}})}>\delta_{\text{th}}$. Here, $v_{\text{D}}^{\max}$ is used as the flight exploration rate of the agent. The final deployment is
	\begin{equation}
		\mathbf{Q}=\{\mathbf{q}_b^{(T_{\text{step}})}\mid b\in\mathcal{U}_B,\ g_{\text{gate},b}^{(T_{\text{step}})}>\delta_{\text{th}}\}.
		\label{eq:deployment_generation}
	\end{equation}
	
\begin{figure}[!t]
		\centering
		\includegraphics[width=\linewidth]{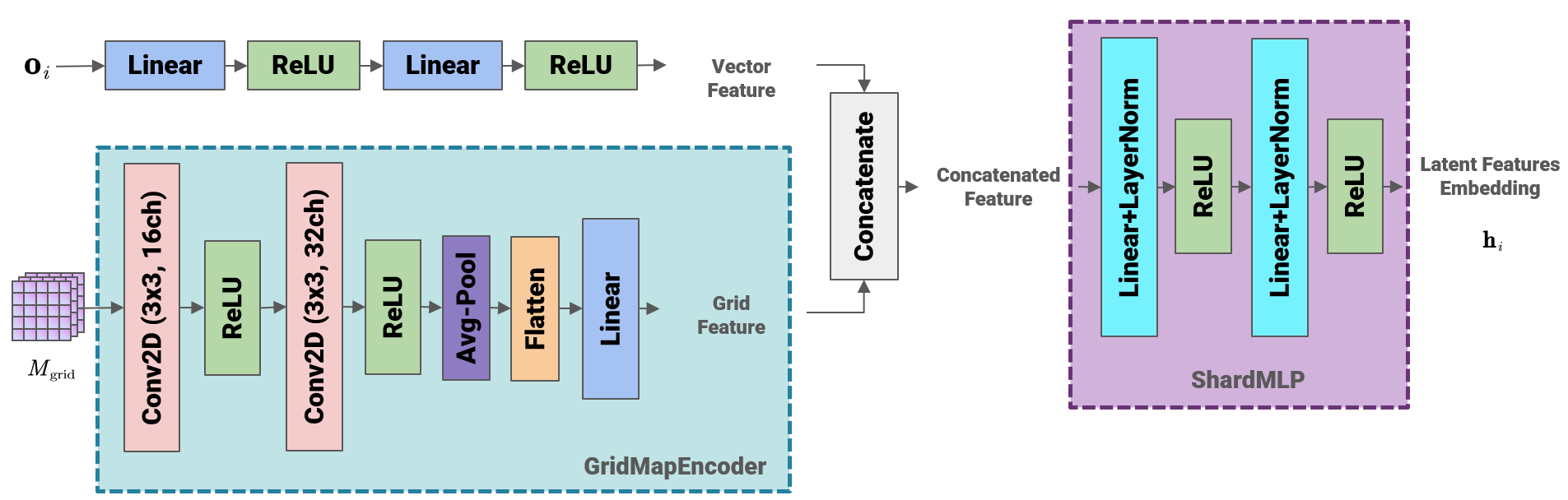}
		\caption{Shared 3D coverage information environment encoder.}
		\label{fig:shard}
	\end{figure}

\begin{figure}[!t]
		\centering
		\includegraphics[width=\linewidth]{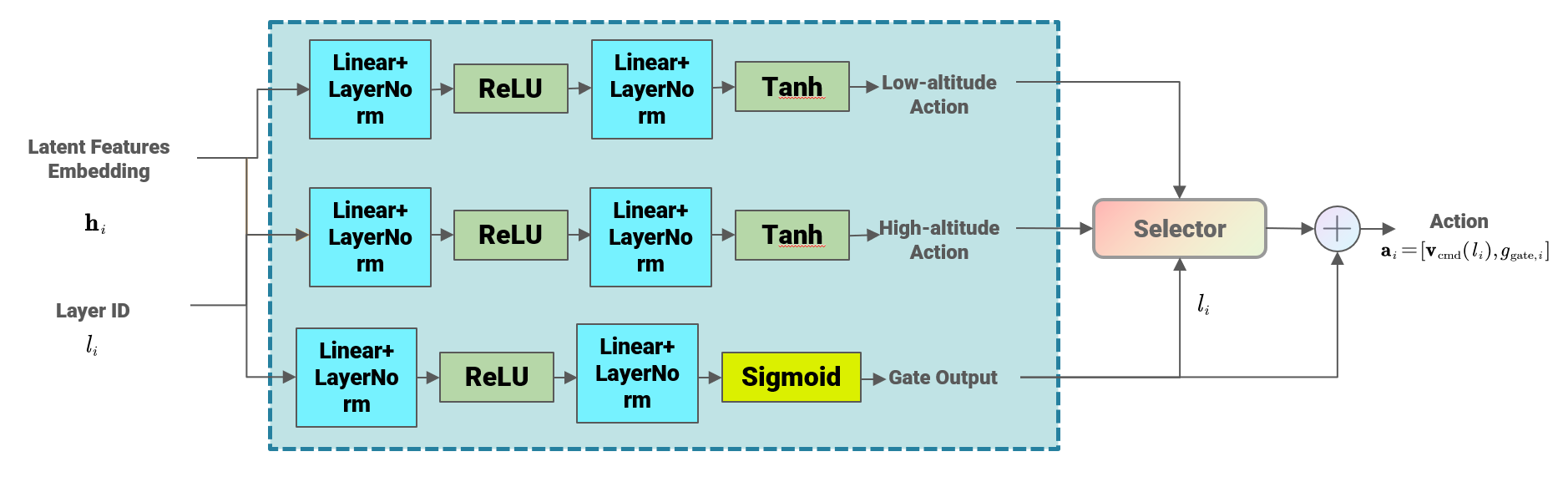}
		\caption{Specific policy output head.}
		\label{fig:role}
	\end{figure}

	\textbf{c) Shared-backbone MATD3 with PER:} Fig.~\ref{fig:matd3_arch} shows the overall training architecture, including the shared encoder, role-aware actor branches, and twin critics under CTDE. The actor is decomposed as
	\begin{equation}
		\mathbf{h}_i=f_{\theta}(\{\mathbf{o}_i,M_{\text{grid}}\}),\qquad
		\mathbf{a}_i=g_{\varphi}(\mathbf{h}_i,l_i),
		\label{eq:shared_backbone_structure}
	\end{equation}
	where $f_{\theta}$ is a shared encoder (Fig.~\ref{fig:shard}) and $g_{\varphi}$ is a role-aware head (Fig.~\ref{fig:role}). This architecture shares spatial features across agents while preserving role-specific control behavior.
	
	Fig.~\ref{fig:shard} and Fig.~\ref{fig:role} illustrate that the shared encoder extracts common spatial features and the role-aware head maps $(\mathbf{h}_i,l_i)$ to motion command and activation gate.
	
	At decision step $\mathfrak{t}$, we denote the joint state and action by $\mathbf{s}_{\mathfrak{t}}=\{M_{\text{grid},\mathfrak{t}},\mathbf{o}_{1,\mathfrak{t}},\ldots,\mathbf{o}_{N_{BS},\mathfrak{t}}\}$ and $\mathbf{a}_{\mathfrak{t}}=\{\mathbf{a}_{1,\mathfrak{t}},\ldots,\mathbf{a}_{N_{BS},\mathfrak{t}}\}$, respectively.
	
	The reward is designed as
	\begin{equation}
		r(\mathbf{s}_{\mathfrak{t}},\mathbf{a}_{\mathfrak{t}})=R_{\text{vol}}+R_{\text{net}}+R_{\text{topo}},
	\end{equation}
	where
	\begin{equation}
		R_{\text{vol}}=\gamma_t C_{\text{term}}+\gamma_v C_{\text{vert}}+\gamma_c C_{\text{corr}},
	\end{equation}
	\begin{equation}
		R_{\text{net}}=\lambda_{\text{conn}}C_{\text{conn}}+\lambda_{\text{cap}}\left(\bar{C}_{\text{syn}}-\bar{C}_{\text{req}}\right),
	\end{equation}
	\begin{equation}
		R_{\text{topo}}=-\sum_{i=1}^{N_{BS}}\left[\sum_{j\in\mathcal{N}_i}\eta_{\text{coll}}\exp\!\left(-\frac{\|\mathbf{q}_i-\mathbf{q}_j\|^2}{\delta_{\text{safe}}^2}\right)+w_h\frac{h_i}{h_{\max}}\right].
	\end{equation}
	Here, $(\gamma_t,\gamma_v,\gamma_c)$ are phase-priority weights in $R_{\text{vol}}$; $(\lambda_{\text{conn}},\lambda_{\text{cap}})$ balance backhaul connectivity and capacity margin in $R_{\text{net}}$; and in $R_{\text{topo}}$, $\mathcal{N}_i$ is the set of $\mathfrak{v}$ nearest neighbors ($\mathfrak{v}=2$), $\eta_{\text{coll}}$ is the collision-penalty scale, $\delta_{\text{safe}}$ is the safety distance, and $w_h$ regularizes altitude (with $h_i$ and $h_{\max}$ denoting current and maximum UAV-BS altitudes). The three terms jointly optimize coverage quality, network robustness/capacity, and geometric safety.
	
	The objective is
	\begin{equation}
		J(\Theta)=\mathbb{E}_{(\mathbf{s}_{\mathfrak{t}},\mathbf{a}_{\mathfrak{t}})\sim\rho^{\pi_{\Theta}}}\left[\sum_{\mathfrak{t}=0}^{T_{\text{step}}}\gamma^{\mathfrak{t}} r\!\left(\mathbf{s}_{\mathfrak{t}},\mathbf{a}_{\mathfrak{t}}\right)\right],
		\label{eq:objective_function}
	\end{equation}
	with $\Theta=\{\theta,\varphi\}$. The shared-encoder gradient is
	\begin{equation}
		\label{eq:shared_gradient}
		\begin{aligned}
			\nabla_{\theta}J(\Theta)
			&\approx \frac{1}{N_{BS}}\sum_{i=1}^{N_{BS}}
			\mathbb{E}_{(\mathbf{s}_{\mathfrak{t}},\mathbf{a}_{\mathfrak{t}})\sim\mathcal{D}}\Big[
			\nabla_{\mathbf{a}_i}Q_{\phi}(\mathbf{s}_{\mathfrak{t}},\mathbf{a}_{\mathfrak{t}}) \\
			&\hspace{4.5em}\cdot\nabla_{\mathbf{h}_i}g_{\varphi}(\mathbf{h}_i,l_i)
			\cdot\nabla_{\theta}f_{\theta}(\{\mathbf{o}_{i,\mathfrak{t}},M_{\text{grid},\mathfrak{t}}\})
			\Big].
		\end{aligned}
	\end{equation}
	
	To reduce overestimation bias, we use twin target critics. Here, $\mathfrak{j}\in\mathcal{B}$ is the sampled-transition index and $\mathfrak{k}\in\{1,2\}$ is the critic index:
	\begin{equation}
		y_{\mathfrak{j}}=r_{\mathfrak{j}}+\gamma\min_{\mathfrak{k}\in\{1,2\}}Q_{\phi_{\mathfrak{k}}'}(\mathbf{s}'_{\mathfrak{j}},\mathbf{a}'_{\mathfrak{j}}),
		\label{eq:matd3_target_update}
	\end{equation}
	and we optimize critics by PER-weighted loss
	\begin{equation}
		\mathcal{L}(\phi_{\mathfrak{k}})=\frac{1}{|\mathcal{B}|}\sum_{\mathfrak{j}\in\mathcal{B}}w_{\mathfrak{j}}\left(y_{\mathfrak{j}}-Q_{\phi_{\mathfrak{k}}}(\mathbf{s}_{\mathfrak{j}},\mathbf{a}_{\mathfrak{j}})\right)^2.
		\label{eq:critic_loss_weighted}
	\end{equation}
	where $w_{\mathfrak{j}}$ is the importance-sampling weight. The full training loop is summarized in Algorithm~\ref{alg:shared_backbone_matd3}.
	
	\begin{algorithm}[htb]
		\caption{3D Coverage-Aware Multi-Agent Deployment via MATD3 with PER}
		\label{alg:shared_backbone_matd3}
		\begin{algorithmic}[1]
			\Require Episodes $E_{\max}$, horizon $T_{\text{step}}$ (max MARL decision steps), number of UAV-BSs $N_{BS}$
			\State Initialize $f_{\theta}$, $g_{\varphi}$, critics $Q_{\phi_1},Q_{\phi_2}$, target networks, and PER buffer $\mathcal{D}$
			\For{$e=1$ to $E_{\max}$}
			\State Reset environment; obtain $M_{\text{grid},0}$ and $\{\mathbf{o}_{i,0}\}_{i=1}^{N_{BS}}$
			\For{$\mathfrak{t}=1$ to $T_{\text{step}}$}
			\For{each agent $i$}
			\State $\mathbf{h}_i\leftarrow f_{\theta}(\{\mathbf{o}_{i,\mathfrak{t}},M_{\text{grid},\mathfrak{t}}\})$
			\State $\mathbf{a}_i\leftarrow g_{\varphi}(\mathbf{h}_i,l_i)+\mathcal{N}(0,\sigma^2)$
			\EndFor
			\State Execute joint action $\mathbf{a}_{\mathfrak{t}}$; observe $r_{\mathfrak{t}},\mathbf{s}_{\mathfrak{t}+1}$; store $(\mathbf{s}_{\mathfrak{t}},\mathbf{a}_{\mathfrak{t}},r_{\mathfrak{t}},\mathbf{s}_{\mathfrak{t}+1})$ in $\mathcal{D}$
			\If{update step}
			\State Sample mini-batch $\mathcal{B}\subset\mathcal{D}$ with PER; compute weights $w_{\mathfrak{j}}$
			\State Compute targets $y_{\mathfrak{j}}$ by Eq.~(\ref{eq:matd3_target_update})
			\State Update $Q_{\phi_1},Q_{\phi_2}$ via Eq.~(\ref{eq:critic_loss_weighted})
			\If{$\mathfrak{t}\bmod d=0$}
			\State Update actor parameters $\Theta=\{\theta,\varphi\}$ using deterministic policy gradient
			\State Soft-update all target networks
			\EndIf
			\EndIf
			\State $\mathbf{s}_{\mathfrak{t}}\leftarrow\mathbf{s}_{\mathfrak{t}+1}$
			\EndFor
			\EndFor
		\end{algorithmic}
	\end{algorithm}
	
	\subsubsection{3D communication-aware A* planner for DPN Trajectory Refinement}
	\label{subsec:algorithm_astar}
	
	\begin{algorithm}[!tb]
		\caption{3D Communication-Aware A* Planner}
		\label{alg:astar_revised}
		\small
		\begin{algorithmic}[1]
			\Require Start cell-node $\xi_s\in\mathcal{X}_{3D}$, goal cell-node $\xi_g\in\mathcal{X}_{3D}$, deployment set $\mathbf{Q}^*$
			\State Initialize OpenSet $\gets\{\xi_s\}$, ClosedSet $\gets\emptyset$
			\State Initialize path-cost-to-come $g(\xi_s)\gets 0$, $f(\xi_s)\gets h(\xi_s)$, and Parent$(\xi_s)=\varnothing$
			\For{all $\xi\in\mathcal{X}_{3D}\setminus\{\xi_s\}$}
			\State $g(\xi)\gets\infty$, $f(\xi)\gets\infty$
			\EndFor
			\While{OpenSet is not empty}
			\State $\xi_{\text{curr}}\gets\arg\min_{\xi\in\text{OpenSet}} f(\xi)$
			\If{$\xi_{\text{curr}}=\xi_g$}
			\State \textbf{break}
			\EndIf
			\State Move $\xi_{\text{curr}}$ from OpenSet to ClosedSet
			\For{each neighbor $\xi_{\text{next}}\in\mathcal{N}_{3D}(\xi_{\text{curr}})$}
			\If{$\xi_{\text{next}}\in\text{ClosedSet}$}
			\State \textbf{continue}
			\EndIf
			\State Compute one-step cost by Eq.~(\ref{eq:edge_cost_astar}) at $\xi_{\text{next}}$:
			\State \hspace{1.2em}$S(\xi_{\text{next}})\gets\max_{b\in\mathbf{Q}^*}\mathrm{SINR}(\xi_{\text{next}},b)$
			\State \hspace{1.2em}\textbf{if} $S(\xi_{\text{next}})<\Gamma_{\text{ctrl}}$ \textbf{then continue} \Comment{equivalent to $\Psi=\infty$}
			\State \hspace{1.2em}$c_{\text{step}}\gets\omega_e\|\xi_{\text{next}}-\xi_{\text{curr}}\|_2+\lambda_{\text{out}}\Psi\!\left(S(\xi_{\text{next}})\right)$
			\State $g_{\text{new}}\gets g(\xi_{\text{curr}})+c_{\text{step}}$
			\If{$g_{\text{new}}<g(\xi_{\text{next}})$}
			\State Parent$(\xi_{\text{next}})\gets \xi_{\text{curr}}$
			\State $g(\xi_{\text{next}})\gets g_{\text{new}}$
			\State $f(\xi_{\text{next}})\gets g(\xi_{\text{next}})+h(\xi_{\text{next}})$  \Comment{A* expansion score}
			\State Insert $\xi_{\text{next}}$ into OpenSet
			\EndIf
			\EndFor
			\EndWhile
			\State Reconstruct path $\mathcal{P}_u^*$ by backtracking Parent$(\cdot)$ from $\xi_g$
			\State \Return $\mathcal{P}_u^*$
		\end{algorithmic}
	\end{algorithm}
	
	Given the deployment $\mathbf{Q}^*$ from Stage Two-step one and the fixed task order from Stage One, we solve for the trajectory variable $\boldsymbol{\mathcal{P}}$ in~(\ref{eq:LL_final}). To avoid conflict with the backhaul graph $\mathcal{G}_n$, we denote the A* search graph by $\mathbb{G}_{3D}=(\mathcal{X}_{3D},\mathcal{A}_{3D})$. The horizontal plane is discretized using the same parameter $K$, and each cell is expanded across altitude layers in $[h_{\min},h_{\max}]$. Each $\xi\in\mathcal{X}_{3D}$ is a feasible 3D cell-node, and each $(\xi,\xi')\in\mathcal{A}_{3D}$ is a feasible one-step transition. The neighborhood is
	\begin{equation}
		\mathcal{N}_{3D}(\xi)=\{\xi'\in\mathcal{X}_{3D}\mid (\xi,\xi')\in\mathcal{A}_{3D}\},
		\label{eq:nbr_def}
	\end{equation}
	which is the one-step reachable set from $\xi$ in $\mathbb{G}_{3D}$.
	
	Consistent with the DPN notation in Section~\ref{set:DUAV}, let $\mathcal{P}_u$ denote UAV-$u$'s path and let its cell-node representation be $\widehat{\mathcal{P}}_u=\{\xi_0,\xi_1,\ldots,\xi_{N_u}\}$, where $N_u$ is the number of one-step transitions, $\xi_0=\xi_s$, and $\xi_{N_u}=\xi_g$. We define
	\begin{equation}
		J_{\text{traj}}(u)=\sum_{\mathfrak{m}=0}^{N_u-1}\left[\omega_e\|\xi_{\mathfrak{m}+1}-\xi_{\mathfrak{m}}\|_2+\lambda_{\text{out}}\,\Psi\big(S(\xi_{\mathfrak{m}+1})\big)\right],
		\label{eq:astar_obj}
	\end{equation}
	where $\mathfrak{m}$ indexes consecutive transitions along $\widehat{\mathcal{P}}_u$, $\omega_e$ weights the motion cost, and $\lambda_{\text{out}}$ weights the communication risk. The best available C2 quality at cell-node $\xi$ is
	\begin{equation}
		S(\xi)=\max_{b\in\mathbf{Q}^*}\mathrm{SINR}(\xi,b).
	\end{equation}
	
	\begin{algorithm}[t]
		\caption{Hierarchical HDNF}
		\label{alg:hierarchical_final}
		\begin{algorithmic}[1]
			\Require Task set $\mathcal{T}$; delivery-UAV set $\mathcal{U}_D$; grid map $M_{\text{grid}}$
			\Require Backhaul SINR threshold $\Gamma_{\text{bh}}$; C2 SINR threshold $\Gamma_{\text{ctrl}}$; pruning tolerance $\epsilon$
			\State Load pre-trained 3D-CASB-MATD3 policy $\pi_\Theta$ with parameters $\Theta=\{\theta,\varphi\}$
			
			\State \textit{// Stage One: Task Assignment and Ordering}
			\State $R_{\text{task}} \leftarrow \text{Heuristic\_VRPTW}(\mathcal{T},\mathcal{U}_D,\omega_e,\omega_{\text{wait}})$
			\If{$R_{\text{task}}=\text{infeasible}$}
			\State \textbf{Return} infeasible (Stage One failed)
			\EndIf
			\State $(\mathbf{X}^*,\{\mathcal{T}_u^*\}) \leftarrow R_{\text{task}}$
			
			\State \textit{// Stage Two, Sub-step 1: ECSN Deployment via 3D-CASB-MATD3}
			\State Construct dual-stream state $\mathbf{S}\leftarrow \text{BuildState}(\{\mathcal{T}_u^*\},M_{\text{grid}})$
			\State $\mathbf{Q}_{\text{init}}\leftarrow \text{Policy\_Inference}(\mathbf{S},\pi_\Theta)$
			\State $\mathbf{Q}^*\leftarrow \mathbf{Q}_{\text{init}}$
			\For{each UAV-BS $b$ in $\mathbf{Q}_{\text{init}}$ (ascending coverage contribution)}
			\State $\mathbf{Q}_{\text{temp}}\leftarrow \mathbf{Q}^*\setminus\{b\}$
			\If{$C_{\text{conn}}(\mathbf{Q}_{\text{temp}})=1$ \textbf{and} $\bar{C}_{\text{syn}}(\mathbf{Q}_{\text{temp}})\ge \bar{C}_{\text{syn}}(\mathbf{Q}_{\text{init}})-\epsilon$}
			\State $\mathbf{Q}^*\leftarrow \mathbf{Q}_{\text{temp}}$ \Comment{Prune while preserving robustness target}
			\EndIf
			\EndFor
			
			\State \textit{// Stage Two, Sub-step 2: DPN Trajectory Refinement}
			\State $\boldsymbol{\mathcal{P}}^*\leftarrow \emptyset$
			\For{each delivery UAV $u\in\mathcal{U}_D$}
			\State $\mathcal{P}_u^*\leftarrow \emptyset$
			\If{$|\mathcal{T}_u^*|=0$}
			\State $\mathcal{P}_u^*\leftarrow\{\mathbf{p}_{\text{center}}\}$
			\Else
			\State Build waypoint sequence $\mathcal{W}_u=[\mathbf{p}_{\text{center}},\boldsymbol{\ell}_{t_{u,1}},\ldots,\boldsymbol{\ell}_{t_{u,|\mathcal{T}_u^*|}},\mathbf{p}_{\text{center}}]$
			\For{$r=1$ to $|\mathcal{W}_u|-1$}
			\State $(\xi_s,\xi_g)\leftarrow\text{MapTo3DCellNodes}(\mathcal{W}_u[r],\mathcal{W}_u[r+1])$
			\State $\mathcal{P}_{u,r}^*\leftarrow \text{CommAware\_A*}(\xi_s,\xi_g,\mathbf{Q}^*)$
			\State $\mathcal{P}_u^*\leftarrow \text{Concat}(\mathcal{P}_u^*,\mathcal{P}_{u,r}^*)$
			\EndFor
			\EndIf
			\State $\boldsymbol{\mathcal{P}}^*\leftarrow \boldsymbol{\mathcal{P}}^*\cup\{\mathcal{P}_u^*\}$
			\EndFor
			
			\State \textbf{Return} joint solution $(\mathbf{X}^*,\mathbf{Q}^*,\boldsymbol{\mathcal{P}}^*)$
		\end{algorithmic}
	\end{algorithm}
	
	The communication penalty is
	\begin{equation}
		\Psi(S)=
		\begin{cases}
			\infty, & S<\Gamma_{\text{ctrl}},\\
			0, &
			\begin{aligned}[t]
				&S\ge\Gamma_{\text{ctrl}},\\
				&S_{\max}\le\Gamma_{\text{ctrl}}+\epsilon_{\Psi},
			\end{aligned}\\
			\max\!\left(0,\frac{S_{\max}-S}{S_{\max}-\Gamma_{\text{ctrl}}}\right),
			&
			\begin{aligned}[t]
				&S\ge\Gamma_{\text{ctrl}},\\
				&S_{\max}>\Gamma_{\text{ctrl}}+\epsilon_{\Psi},
			\end{aligned}
		\end{cases}
		\label{eq:penalty_func}
	\end{equation}
	where $S_{\max}=\max_{\xi\in\mathcal{X}_{3D},\,b\in\mathbf{Q}^*}\mathrm{SINR}(\xi,b)$ denotes the maximum attainable SINR under the current deployment and search space, and $\epsilon_{\Psi}>0$ is a small safeguard. This penalty rejects outage nodes and decreases as feasible SINR increases.
	
	For path-planner search, the heuristic function is
	\begin{equation}
		h(\xi)=\omega_e\|\xi-\xi_g\|_2,
		\label{eq:heuristic}
	\end{equation}
	To connect with the Stage Two subproblem in Eq.~(\ref{eq:LL_final}), we define the edge cost induced by Eq.~(\ref{eq:astar_obj}) as
	\begin{equation}
		c(\xi,\xi')=\omega_e\|\xi'-\xi\|_2+\lambda_{\text{out}}\,\Psi\!\left(S(\xi')\right),\quad (\xi,\xi')\in\mathcal{A}_{3D}.
		\label{eq:edge_cost_astar}
	\end{equation}
	Then the dynamic recursion of the realized path cost is
	\begin{equation}
		g(\xi')=\min_{\xi:(\xi,\xi')\in\mathcal{A}_{3D}}\left[g(\xi)+c(\xi,\xi')\right],\quad g(\xi_s)=0,
		\label{eq:g_recurrence}
	\end{equation}
	and the A* evaluation score is
	\begin{equation}
		f(\xi)=g(\xi)+h(\xi).
		\label{eq:f_def}
	\end{equation}
	
	Algorithm~\ref{alg:astar_revised} summarizes the planner. Thus, Stage Two-(2) solves the trajectory subproblem in Eq.~(\ref{eq:LL_final}) by minimizing $J_{\text{traj}}$ through A* expansion with $f(\xi)=g(\xi)+h(\xi)$.
	
	\subsection{Algorithm Overview}
	\label{subsec:algorithm_overview}
	
	Algorithm~\ref{alg:hierarchical_final} summarizes the complete HDNF inference pipeline and is consistent with the stage decomposition in Section~\ref{sec:method}. The pipeline first executes Stage One (task assignment and ordering). If Stage One is infeasible, the procedure terminates and returns infeasible; otherwise, it outputs $(\mathbf{X}^*,\{\mathcal{T}_u^*\})$.
	
	Stage Two is then executed in two sequential sub-steps corresponding to the two subsubsections under Stage Two. First, the ECSN-deployment sub-step uses the ordered tasks together with the communication-demand cues in $M_{\text{grid}}$ to infer an initial deployment $\mathbf{Q}_{\text{init}}$. A pruning step then refines $\mathbf{Q}_{\text{init}}$ into $\mathbf{Q}^*$ while preserving the target robustness level ($C_{\text{conn}}=1$) and maintaining synchronized C2 performance within tolerance $\epsilon$. Second, the DPN-trajectory-refinement sub-step is performed based on $\mathbf{Q}^*$. For each delivery UAV, if no task is assigned, a depot-only trajectory is returned; otherwise, the ordered mission waypoints are decomposed into consecutive waypoint pairs, each pair is solved by the communication-aware A* planner, and the resulting local paths are concatenated to form the final 3D route. Collecting all refined trajectories yields $\boldsymbol{\mathcal{P}}^*$ and completes the coordinated output $(\mathbf{X}^*,\mathbf{Q}^*,\boldsymbol{\mathcal{P}}^*)$.
	
	\section{Experiments and Results}
	\label{sec:experiments}
	This section evaluates the advantages of the proposed HDNF in communication-coverage quality and training efficiency, and further highlights the importance of the multi-layer C2 service model. We compare against the following schemes:
	\begin{enumerate}
		\item \textbf{Proposed (HDNF):} ECSN deployment is optimized using the 3D-CASB MATD3 with PER algorithm integrated with the multi-layer C2 service model, and DPN routing is solved by the proposed 3D communication-aware A* planner.
		\item \textbf{MATD3-2D (2D-Only):} This baseline implements the DRL-based deployment strategy proposed in \cite{UAV_BS_4}. It utilizes the specified deep neural network (DNN) architecture to perceive the spatial distribution of ground tasks via grid-based heatmaps, guiding the UAV-BS placement based on 2D ground-centric demand. In our framework, this method is restricted to conventional 2D coverage, and DPN routing is solved by our 3D communication-aware A* planner.
		\item \textbf{Grid Deployment:} ECSN deployment uses a static grid-based deployment strategy, while DPN routing is solved by the same 3D communication-aware A* planner.
		\item \textbf{Ablation w/o PER:} ECSN deployment uses the proposed framework but removes the PER mechanism.
		\item \textbf{Ablation w/o Shared Backbone:} ECSN deployment uses the proposed framework but removes the Shared Backbone architecture, so each agent extracts environmental features independently.
	\end{enumerate}
	All simulation parameters and hyperparameter settings are summarized in Table~\ref{tab:sim_params}. For reproducibility, task payloads are sampled i.i.d. as $w_t\sim\mathcal{U}[0.5,1.5]$ kg and then fixed across all compared schemes under the same random seed for each scenario.
	\begin{table}[!t]
		\caption{Simulation Parameters and Hyperparameter Settings}
		\label{tab:sim_params}
		\centering
		\scriptsize
		\setlength{\tabcolsep}{3pt}
		\renewcommand{\arraystretch}{1.2}
		
		\begin{tabularx}{\linewidth}{@{} >{\raggedright\arraybackslash}X c c @{}}
			\toprule
			\textbf{Parameter} & \textbf{Symbol} & \textbf{Value} \\
			\midrule
			
			\multicolumn{3}{@{}l}{\textit{\textbf{Environment \& Kinematics}}} \\
			Target Area Dimension & $\Omega$ & $\{3000,3500, \dots, 5000\}^2$ m$^2$ \\
			Central Depot Location & $\mathbf{p}_{\text{center}}$ & $(200, 100, 0)$ \\
			Number of Delivery UAVs & $U_D$ & $7$ \\
			Maximum Number of UAV-BSs & $N_{BS}$ & $15, 20, 25, 30, 35$ \\
			Delivery-UAV Cruise Altitude & $h_{\text{cruise}}$ & $100$ m \\
			UAV-BS Altitude Range & $[h_{\min}, h_{\max}]$ & $[30, 200]$ m \\
			Each UAV Maximum Flight Speed & $v_{\text{D}}^{\max}$ & $20$ m/s \\
			Number of Delivery Tasks & $T_{\text{task}}$ & $30$ \\
			Maximum Payload Capacity & $W_u^{\max}$ & $15$ kg \\
			Delivery UAV Battery Capacity & $E_u^{\max}$ & $10000$ kJ \\
			Flight-Energy Consumption Coefficient & $\eta$ & $500$ J/(m$\cdot$kg) \\
			\addlinespace[1pt]
			\multicolumn{3}{@{}l}{\textit{\textbf{Multi-Layer C2 Sampling}}} \\
			Vertical Sampling Start Altitude & $h_{\min}^{(v)}$ & $0$ m \\
			Vertical Sampling Interval & $\Delta h$ & $10$ m \\
			Number of Vertical Sampling Intervals & $M_v$ & $15$ \\
			Corridor Sampling Density Factor & $I_t$ & $20$ \\
			Grid Discretization Resolution & $K \times K$ & $100 \times 100$ \\
			Collision Avoidance Radius & $\delta_{\text{safe}}$ & $250$ m \\
			Infeasibility Penalty Constant & $\mathcal{M}_{\infty}$ & $10^{10}$ \\
			
			\midrule
			\multicolumn{3}{@{}l}{\textit{\textbf{Communication Model}}} \\
			Carrier Frequency & $f_c$ & $2.4$ GHz \\
			Transmit Power & $P_t$ & $23$ dBm \\
			Noise Power Spectral Density & $N_0$ & $-174$ dBm/Hz \\
			System Channel Bandwidth & $B$ & $10$ MHz \\
			C2 Control-Link SINR Threshold (dB) & $\Gamma_{\text{ctrl}}$ & $14$ dB \\
			Backhaul-Link SINR Threshold (dB) & $\Gamma_{\text{bh}}$ & $12$ dB \\
			Maximum Spectral Efficiency & $C_{\max}$ & $8$ bps/Hz \\
			Propagation Speed & $c_0$ & $3 \times 10^8$ m/s \\
			Additional Path Loss (LoS/NLoS) & $\eta_{\text{LoS}}, \eta_{\text{NLoS}}$ & $1.0, 20.0$ \\
			Urban Environment Parameters from \cite{se} & $\alpha, \beta$ & $9.61, 0.16$ \\
			
			\midrule
			\multicolumn{3}{@{}l}{\textit{\textbf{3D CASB-MATD3 with PER Hyperparameters}}} \\
			Actor Learning Rate & $\alpha_{\theta}$ & $1 \times 10^{-4}$ \\
			Critic Learning Rate & $\alpha_{\phi}$ & $5 \times 10^{-4}$ \\
			Replay Buffer Size & $|\mathcal{D}|$ & $1 \times 10^6$ \\
			Mini-Batch Size & $|\mathcal{B}|$ & $128$ \\
			Discount Factor & $\gamma$ & $0.99$ \\
			Target Network Soft Update Rate & $\tau_{\text{soft}}$ & $0.005$ \\
			Exploration Noise Std. Dev. & $\sigma$ & $0.2$ \\
			Policy Update Frequency & $d$ & $2$ \\
			Deployment-gate Score & $g_{\text{gate}}$& $0.5$\\
			
			\midrule
			\multicolumn{3}{@{}l}{\textit{\textbf{Objective \& Reward Coefficients}}} \\
			Flight Energy Objective Weight & $\omega_e$ & $1.0$ \\
			Waiting Penalty Weight & $\omega_{\text{wait}}$ & $0.5$ \\
			C2 Service Model Weights & $(\omega_t, \omega_v, \omega_c)$ & $0.3, 0.3, 0.4$ \\
			Deployment Cost Objective Weight & $\omega_d$ & $0.5$ \\
			C2 Coverage Reward Weights & $(\gamma_t, \gamma_v, \gamma_c)$ & $2.0, 2.0, 4.0$ \\
			Backhaul Connectivity Reward Weight & $\lambda_{\text{conn}}$ & $5.0$ \\
			Normalized Capacity Incentive & $\lambda_{\text{cap}}$ & $1.0$ \\
			Required Synchronized-Capacity Baseline & $\bar{C}_{\text{req}}$ & Mission-dependent \\
			Desired Robustness Level & $\lambda_{\text{req}}$ & $0.5$ \\
			Collision-Penalty Scaling Factor & $\eta_{\text{coll}}$ & $50.0$ \\
			Altitude-Variation Penalty Weight & $w_h$ & $0.5$ \\
			C2 Outage Penalty Weight & $\lambda_{\text{out}}$ & $1000.0$ \\
			\bottomrule
		\end{tabularx}
	\end{table}
	
	\subsection{Training Reward Convergence and Training Efficiency Analysis}
	\begin{figure}[h]
		\centering
		\includegraphics[width=\linewidth]{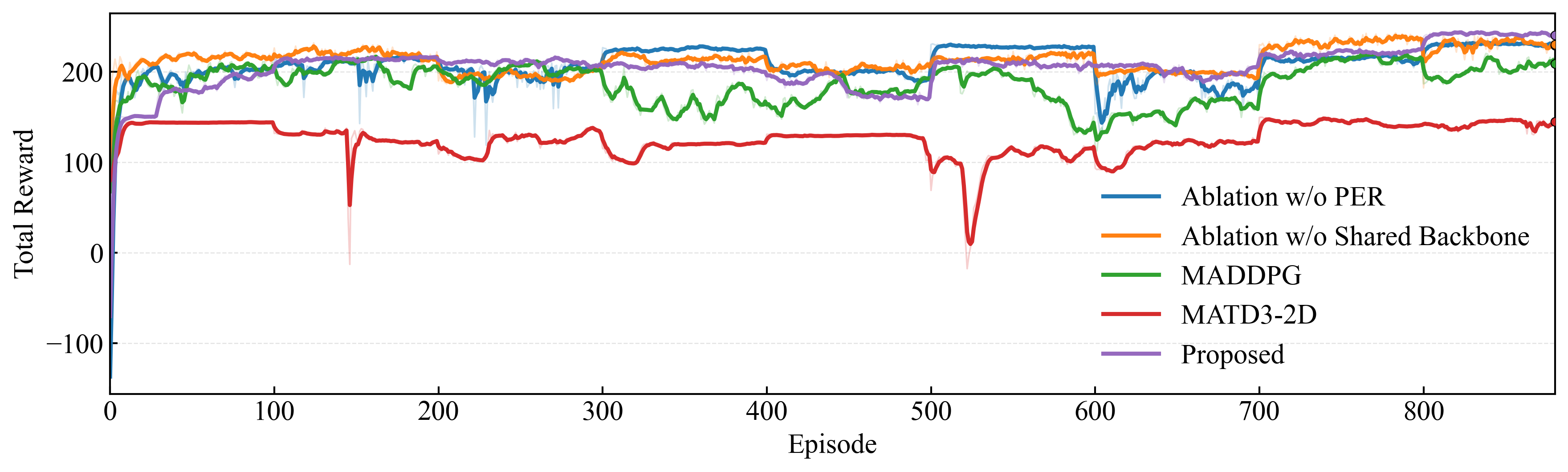}
		\caption{Total reward convergence.}
		\label{fig:reward_convergence}
	\end{figure}
	
	Figure~\ref{fig:reward_convergence} shows total reward trajectories over 900 training episodes. MADDPG exhibits large oscillations and unstable convergence, indicating weak coordination in high-dimensional continuous spaces. MATD3-2D (2D-Only) converges quickly but reaches the lowest asymptotic reward because it ignores 3D aerial communication demand. HDNF converges rapidly and stabilizes at the highest reward level. The two ablations confirm the contributions of both components: removing the Shared Backbone increases late-stage oscillation, while removing PER reduces early exploration efficiency and slows reward accumulation.
	
	\begin{figure}[!t]
		\centering
		\includegraphics[width=0.6\linewidth]{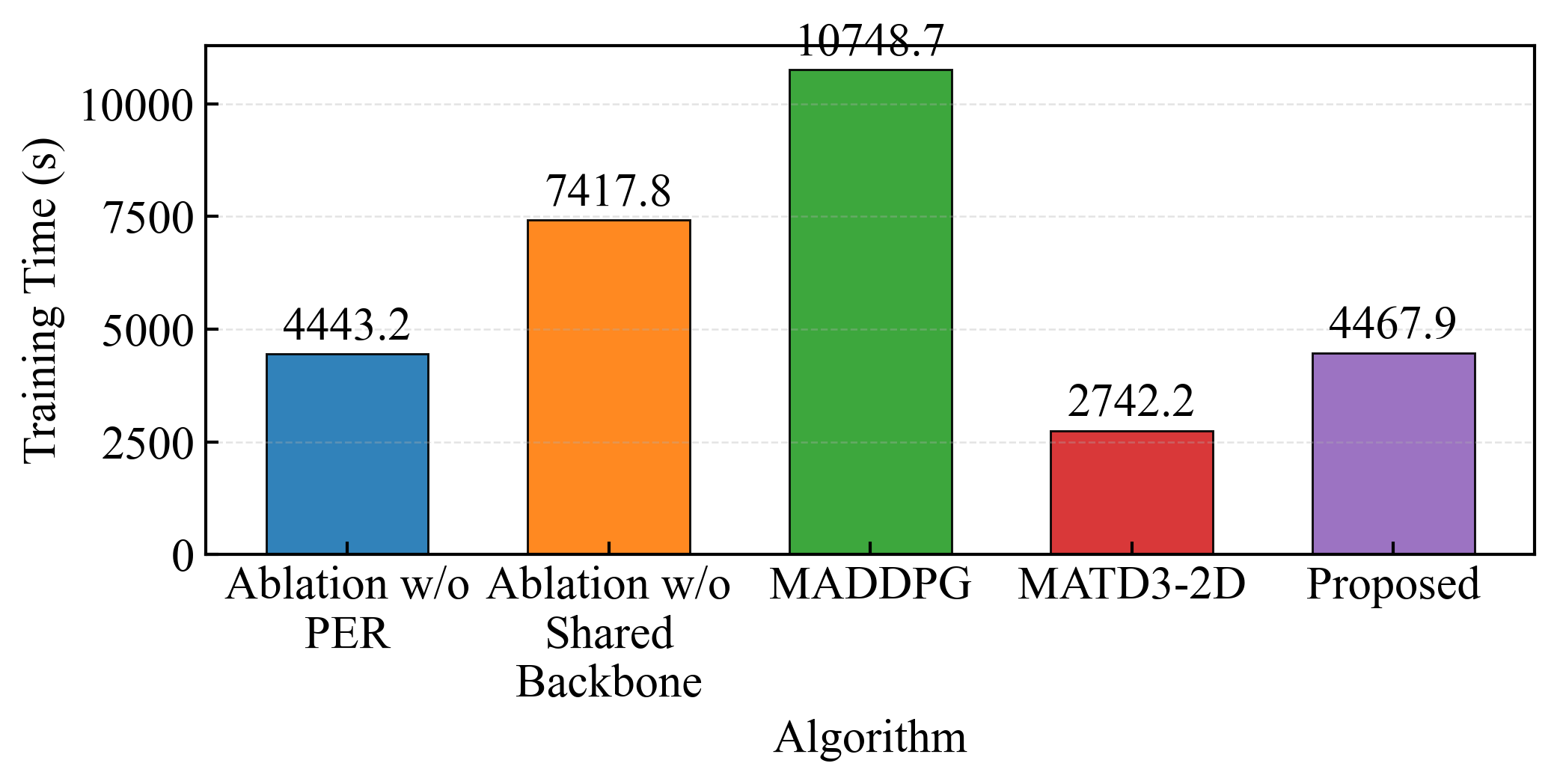}
		\caption{Total training time to reach the target number of episodes.}
		\label{fig:training_time_bar}
	\end{figure}
	
	Figure~\ref{fig:training_time_bar} reports wall-clock training time to reach the same target episode count. HDNF reduces training time by about 58.4\% versus MADDPG and 39.8\% versus Ablation w/o Shared Backbone. Although MATD3-2D has the shortest runtime, this speedup comes from omitting high-dimensional 3D feature extraction and is consistent with its lowest asymptotic reward in Fig.~\ref{fig:reward_convergence}. Overall, the Shared Backbone improves training efficiency while preserving policy quality.
	
	\subsection{Analysis of the Performance of Multi-Layer C2 Service Model}
	\begin{figure}[!t]
		\centering
		\includegraphics[width=\linewidth]{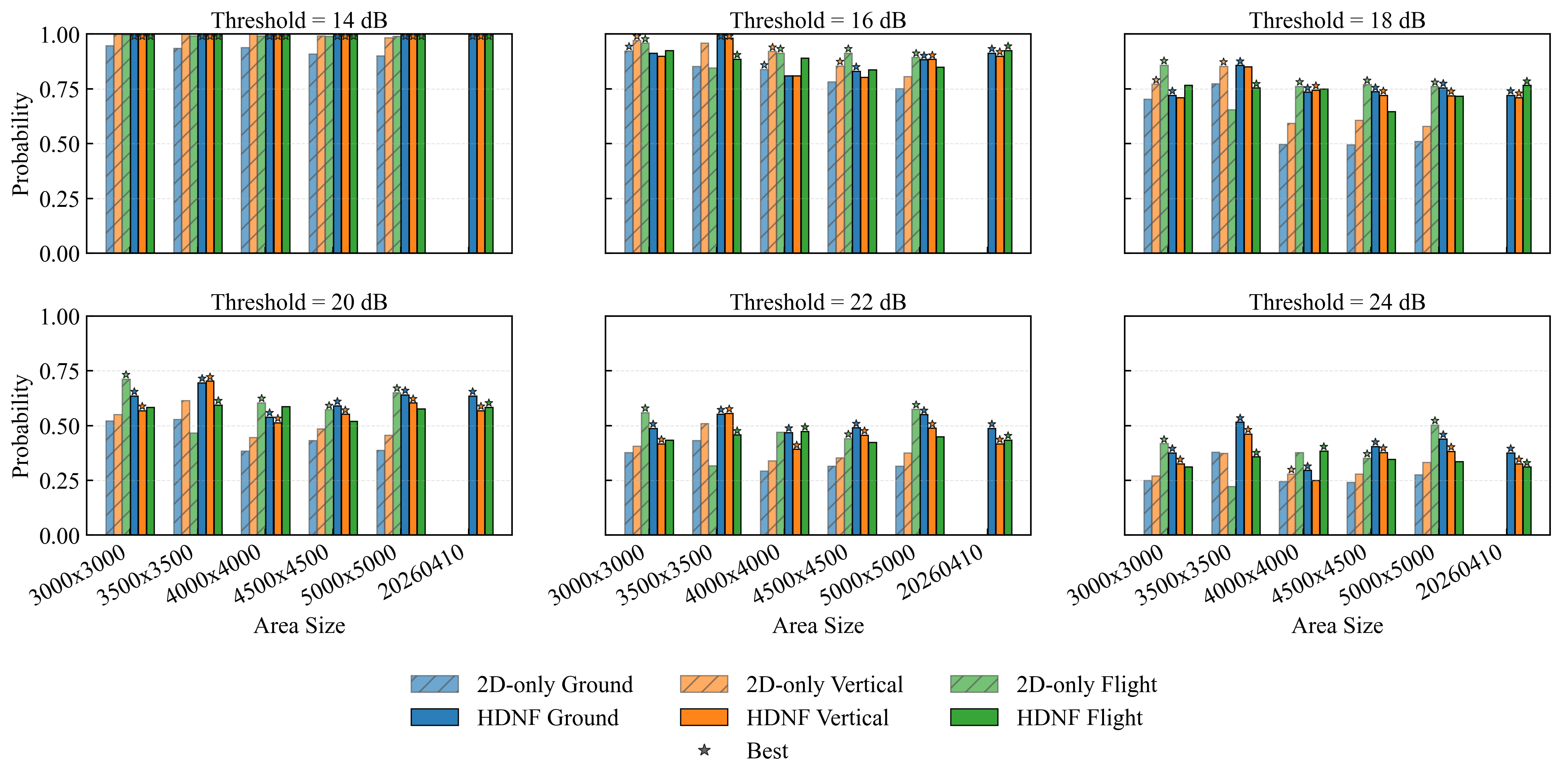}
		\caption{Probability Distribution of C2 Link Quality at Different SINR Requirements.}
		\label{fig:sinr_stage_probability}
	\end{figure}
	
	\begin{figure}[!t]
		\centering
		\includegraphics[width=\linewidth]{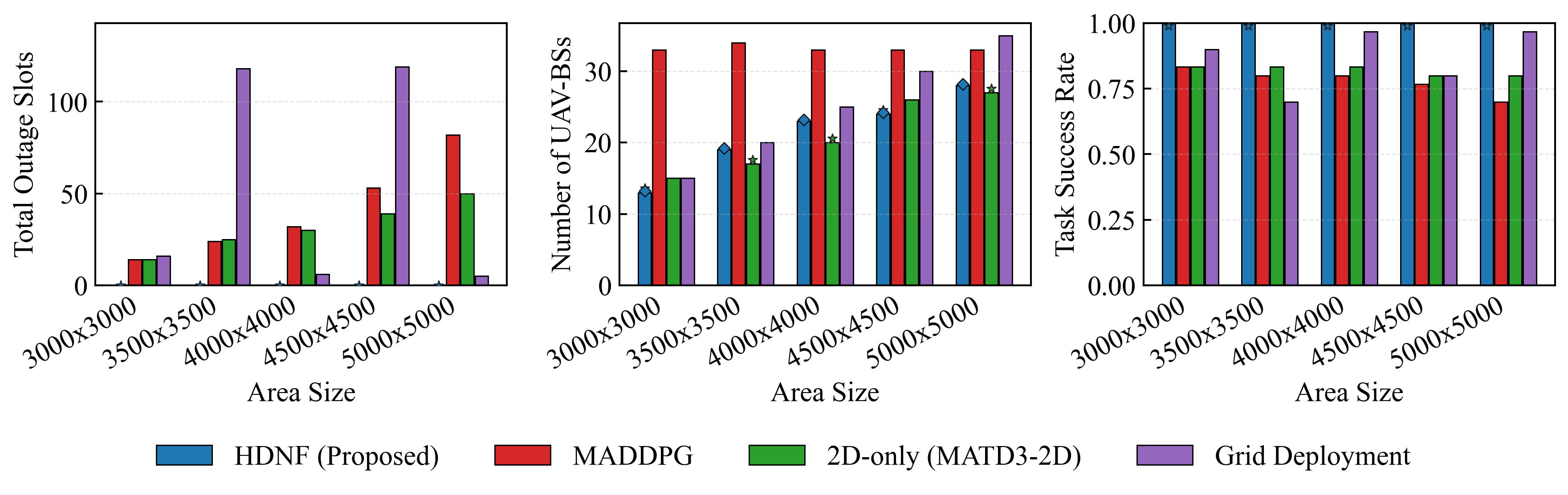}
		\caption{System-level performance metrics under varying disaster area sizes.}
		\label{fig:benchmark_metrics}
	\end{figure}
	
	\begin{figure}[!t]
		\centering
		\includegraphics[width=\linewidth]{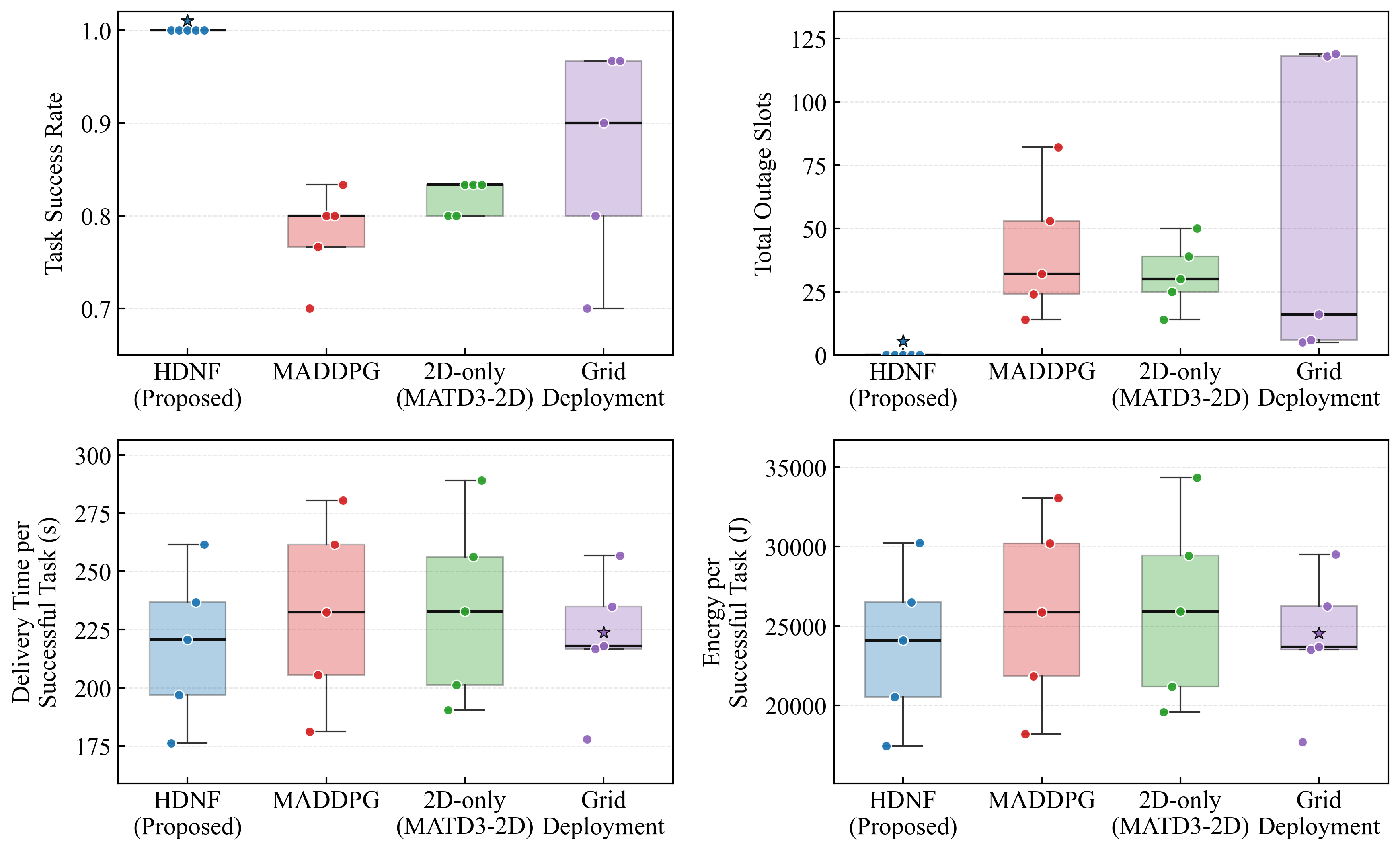}
		\caption{Boxplot-based statistical distributions of key performance indicators under different schemes.}
		\label{fig:benchmark_boxplot}
	\end{figure}
	
	\begin{figure*}[!hbt]
		\centering
		\includegraphics[width=\textwidth]{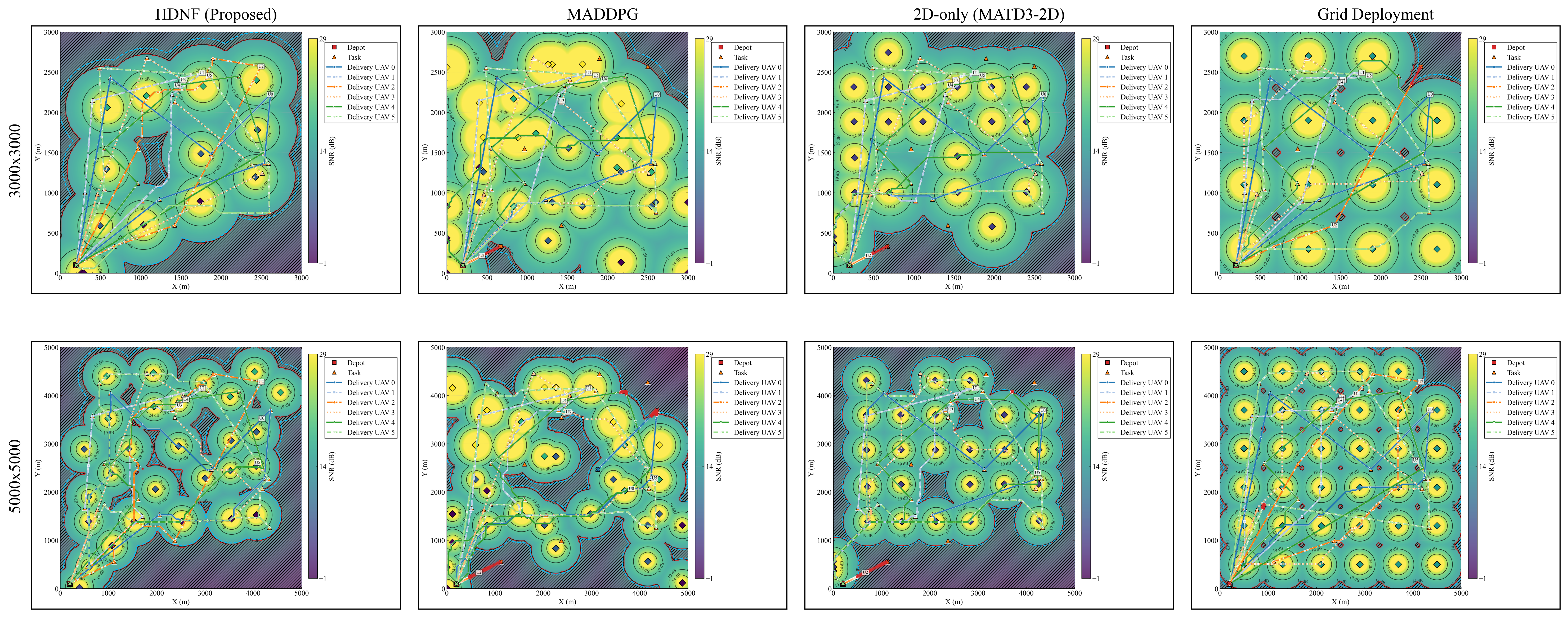}
		\caption{Coverage heatmaps of UAV-BS deployment schemes in representative post-disaster areas.}
		\label{fig:heatmap_panel}
	\end{figure*}
	
	\begin{figure*}[!hbt]
		\centering
		\includegraphics[width=\textwidth]{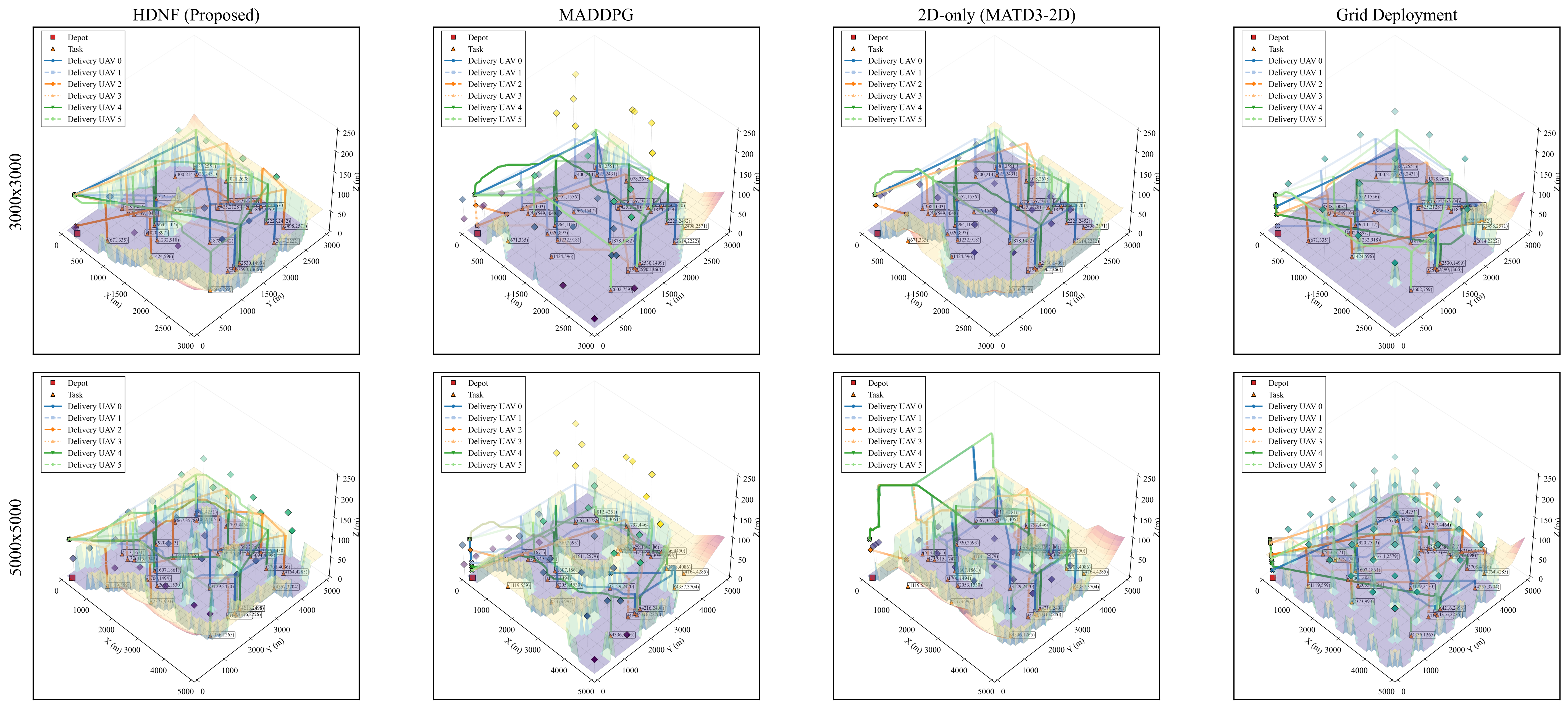}
		\caption{3D delivery-UAV trajectories under different deployment schemes in representative post-disaster areas.}
		\label{fig:isometric_panel}
	\end{figure*}
	
	To evaluate the practical value of the multi-layer C2 service model for ECS and DPN coordination, we compare HDNF with 2D-Only under different disaster scales and SINR requirements. Figure~\ref{fig:sinr_stage_probability} reports C2 satisfaction probabilities from 14 dB to 24 dB for three mission phases: terminal delivery (Ground), vertical takeoff/landing (Vertical), and high-altitude cruise (Flight). Under loose requirements (e.g., 14 dB), both methods provide basic coverage, though 2D-Only already lags behind. Furthermore, this performance evaluation must be contextualized within the strict upper bound of deployable UAV-BSs. As requirements tighten (16 dB to 24 dB), the overall coverage probabilities inevitably decline across all methods due to this rigid UAV-BSs number. Notably, while the 2D-centric baseline may occasionally over-index on a single isolated phase (e.g., ground) at elevated thresholds to yield a locally higher probability, it does so at the catastrophic expense of the remaining 3D trajectory. Conversely, HDNF systematically balances the communication demands across all critical phases, thereby averting the severe numerical collapse observed in the baseline's vertical and flight stages. Considering HDNF guarantees this holistic 3D communication assurance while utilizing up to $20\%$ fewer base stations, it proves that our framework fundamentally maximizes the comprehensive coverage yield per UAV-BS under extreme channel constraints.
	\subsection{System-Level Scalability and Robustness Assessment of Dual-Network Coordination}
	To evaluate system-level scalability, Fig.~\ref{fig:benchmark_metrics} reports total outage slots, deployed UAV-BS count, and task success rate as the disaster area expands from $3000\times3000$ m$^2$ to $5000\times5000$ m$^2$. We adopt a strict failure policy: once a delivery UAV experiences a communication outage, all its remaining assigned tasks are counted as failed.
	
	Analyzing the metrics individually, the first subplot shows that MADDPG and 2D-Only suffer from rapidly increasing outage slots as the area expands, while Grid Deployment experiences severe erratic spikes. Second, regarding UAV-BSs consumption, Grid and MADDPG quickly exhaust the maximum limit of 35 UAV-BSs, whereas HDNF scales efficiently, requiring only 28 UAV-BSs in the largest 5000x5000 m² scenario (a $20\%$ UAV-BS saving). Consequently, the third subplot confirms that HDNF uniquely sustains a perfect $100\%$ task success rate, while MADDPG and 2D-Only degrade below $80\%$. Ultimately, these phenomena occur because HDNF's multi-layer 3D coordination effectively eliminates the spatial coverage mismatches and resource inefficiencies that plague traditional ground-centric or static deployments.
	
	To complement mean-level comparisons, Fig.~\ref{fig:benchmark_boxplot} presents boxplots of task success rate, total outage slots, average successful-task delivery time, and average successful-task energy consumption. HDNF shows the best reliability profile, with success rates tightly concentrated at 1.0 and outage slots near zero, both with low dispersion. MADDPG and 2D-Only have median success rates around 0.8 and notably broader outage distributions, indicating weaker communication robustness. Grid Deployment occasionally yields slightly lower time/energy medians because dense static redundancy can create locally over-covered corridors and near-straight routes for completed tasks. However, this gain depends on substantial infrastructure over-provisioning. HDNF achieves near-comparable efficiency with far fewer UAV-BSs while eliminating outages, demonstrating a stronger balance among deployment cost, mission reliability, and execution efficiency.
	
	\subsection{Coordination Analysis of ECSN and DPN}
	To show how different deployment strategies affect delivery execution, we jointly analyze the 2D coverage heatmaps in Fig.~\ref{fig:heatmap_panel} and the 3D delivery trajectories in Fig.~\ref{fig:isometric_panel}. In Fig.~\ref{fig:heatmap_panel}, the red boundary denotes the coverage threshold boundary, the area outside this boundary corresponds to coverage blind zones, the blue dashed line denotes the backhaul-link SINR threshold line, and the red path denotes a signal-interruption path. Fig.~\ref{fig:heatmap_panel} shows that HDNF maintains effective coverage around task locations in both the $3000\times3000$ m$^2$ and $5000\times5000$ m$^2$ task regions while using fewer UAV-BSs, whereas the other schemes show insufficient coverage as the task region expands. In the $5000\times5000$ m$^2$ case, both MADDPG and 2D-Only leave some task locations uncovered. This difference directly affects the flight trajectories. Because HDNF considers communication requirements in the terminal, vertical, and cruise phases, most delivery UAVs travel between tasks along paths close to straight horizontal routes. By contrast, under MADDPG and 2D-Only, incomplete UAV-BS coverage forces delivery UAVs to detour toward better-covered areas, increasing path irregularity and energy consumption. When task locations are not effectively covered, feasible communication-aware routes may not exist, so some missions cannot be completed. Fig.~\ref{fig:isometric_panel} further shows the altitude behavior of the planned routes. Under the baseline schemes, delivery UAVs often climb or descend above non-task areas to search for better communication locations before continuing toward the destination, which further increases path length and energy consumption. In contrast, under HDNF, delivery UAVs can maintain an almost fixed cruise altitude before reaching the airspace above the target task location and then perform the required vertical descent or ascent near the service point. Overall, the proposed dual-network coordination reduces unnecessary detours and altitude changes, improving delivery efficiency and reliability.
	
	\section{Conclusion}
	\label{sec:conclusion}
	In post-disaster scenarios with damaged ground infrastructure, the safe operation of emergency-delivery UAVs require continuous C2 connectivity. Traditional systems suffer from a mismatch between static ground-centric coverage and the 3D flight requirements of delivery UAVs. To address this issue, we has proposed the HDNF to coordinate the ECSN and DPN. With a multi-layer C2 service model, a 3D coverage-aware MARL algorithm, and a communication-aware A* planner, HDNF eliminate C2 outages in key phases, including takeoff, landing, and cruise. Extensive results have shown that the HDNF can maintain a higher task success rate under strict constraints and reduce the number of required UAV-BSs by up to $20\%$ compared with conventional static deployments, thereby providing an efficient and reliable system-level solution.
	
	\FloatBarrier
	\bibliographystyle{IEEEtran}
	\bibliography{new}
	
\end{document}